\documentclass[aps,prd,showpacs,superscriptaddress,groupedaddress,nofootinbib]{revtex4}

\usepackage{graphicx}
\usepackage{dcolumn}
\usepackage{bm}

\usepackage{inputenc}
\usepackage{float}
\usepackage{amsmath}
\usepackage{amssymb}
\usepackage{mathtools}
\usepackage{siunitx}
\usepackage{amsfonts}
\usepackage{dsfont}
\usepackage{array}
\usepackage{mathrsfs}
\usepackage{pifont}
\usepackage{multirow}
\usepackage{upgreek}
\usepackage{xcolor}
\usepackage{hyperref}
\usepackage{verbatim}
\usepackage{slashed}
\usepackage{cleveref}
\usepackage{ucs}
\usepackage{tikz}
\usepackage{subfigure}
\usepackage{caption}
\usepackage{makecell}
\usepackage{rotating}

\usepackage[normalem]{ulem}

\usetikzlibrary{arrows}
\usetikzlibrary{decorations.markings}

\newcommand{\tab}[1]{Table~\ref{#1}}

\newcommand{\gev}{\,\mathrm{GeV}}

\newcommand{\lbr}[1][6]{\displaybreak[1]\\[#1pt]}

\DeclareMathOperator{\Tr}{Tr}

\newcommand{\MSbar}{\overline{\text{MS}}}

\newcommand{\RI}{\text{RI}}

\begin{document}


\title{Nonperturbative renormalization of asymmetric staple-shaped \\operators in twisted
mass lattice QCD}

\author{Constantia Alexandrou}
\email{alexand@ucy.ac.cy}
\affiliation{Department of Physics, University of Cyprus, P.O. Box 20537, 1678 Nicosia, Cyprus}
\affiliation{Computation-based Science and Technology Research Center, The Cyprus Institute, 20 Kavafi Street, Nicosia 2121, Cyprus}
\author{Simone Bacchio}
\affiliation{Computation-based Science and Technology Research Center, The Cyprus Institute, 20 Kavafi Street, Nicosia 2121, Cyprus}
\author{Krzysztof Cichy}
\affiliation{Faculty of Physics, Adam Mickiewicz University, ul. Uniwersytetu Poznańskiego 2, 61-614 Poznań, Poland}
\author{Martha Constantinou}
\affiliation{Temple University,1925 N. 12th Street, Philadelphia, Pennsylvania 19122-1801, USA}
\author{Xu Feng}
\affiliation{School of Physics and State Key Laboratory of Nuclear Physics and Technology, Peking University, Beijing 100871, China}
\affiliation{Collaborative Innovation Center of Quantum Matter, Beijing 100871, China}
\affiliation{Center for High Energy Physics, Peking University, Beijing 100871, China}
\author{Karl Jansen}
\affiliation{NIC, DESY, Platanenallee 6, D-15738 Zeuthen, Germany}
\author{Chuan Liu}
\affiliation{School of Physics and State Key Laboratory of Nuclear Physics and Technology, Peking University, Beijing 100871, China}
\affiliation{Collaborative Innovation Center of Quantum Matter, Beijing 100871, China}
\affiliation{Center for High Energy Physics, Peking University, Beijing 100871, China}
\author{Aniket Sen}
\email{sen@hiskp.uni-bonn.de}
\affiliation{Helmholtz Institut für Strahlen- und Kernphysik, Rheinische Friedrich-Wilhelms-Universität Bonn, Nussallee 14-16, 53115 Bonn}
\affiliation{Bethe Center for Theoretical Physics, Rheinische Friedrich-Wilhelms-Universität Bonn, 
Nussallee 12, 53115 Bonn, Germany}
\author{Gregoris Spanoudes}
\affiliation{Department of Physics, University of Cyprus, P.O. Box 20537, 1678 Nicosia, Cyprus}
\affiliation{Computation-based Science and Technology Research Center, The Cyprus Institute, 20 Kavafi Street, Nicosia 2121, Cyprus}
\author{Fernanda Steffens}
\affiliation{Helmholtz Institut für Strahlen- und Kernphysik, Rheinische Friedrich-Wilhelms-Universität Bonn, Nussallee 14-16, 53115 Bonn}
\affiliation{Bethe Center for Theoretical Physics, Rheinische Friedrich-Wilhelms-Universität Bonn, 
Nussallee 12, 53115 Bonn, Germany}
\author{Jacopo Tarello}
\affiliation{Department of Physics, University of Cyprus, P.O. Box 20537, 1678 Nicosia, Cyprus}
\affiliation{Computation-based Science and Technology Research Center, The Cyprus Institute, 20 Kavafi Street, Nicosia 2121, Cyprus}

\date{\today}

\begin{abstract}

Staple-shaped Wilson line operators are necessary 
for the study of transverse momentum-dependent parton 
distribution functions (TMDPDFs) in lattice QCD and beyond. 
In this work, we study the renormalization of such operators 
in the general case of an asymmetric staple.
We analyze the mixing pattern of these operators using their 
symmetry properties, where we find that the possible mixing 
is restricted within groups of four operators. We then present 
numerical results using the regularization independent momentum subtraction
(RI/MOM) scheme to study the importance of mixing using one operator in
particular, the $\gamma_0$ operator. Based on these results, 
we consider the short distance ratio (SDR) scheme, which is desirable 
in the absence of mixing. Finally, we investigate a variant of 
the RI/MOM scheme, where the renormalization factors are computed 
at short distances.
\end{abstract}

\maketitle

\section{Introduction}

Collinear parton distribution functions (PDFs) probe  the 
hadron structure from the perspective of the spin and 
longitudinal momentum distributions of the quarks and gluons 
that make up the hadron. 
The vast amount of work to determine these distributions 
over the last five decades, both theoretically and 
experimentally,  has enormously expanded our view of 
the hadron structure, the proton in particular~\cite{Amoroso:2022eow}. 
And yet, these developments are mostly limited to probing 
the  one-dimensional structure of the proton. 
In order to have a wider understanding of the proton 
structure, we also need to understand how the momentum 
and the spin are distributed in the transverse plane. 
For that, we need to measure and compute the generalized
parton distributions and the 
transverse-momentum-dependent PDFs (TMDPDFs), the latter
being the main focus of this present work.
Although there has been an effort to obtain TMDPDFs from 
phenomenological fits to experimental data~\cite{Bacchetta:2017gcc,Bertone:2019nxa,Scimemi:2019cmh,Bury:2022czx,Bacchetta:2022awv,Barry:2023qqh}, 
their accuracy is still far from being 
at the same level as the collinear PDFs. 
This status will change in the coming years with new 
data coming from Jefferson Lab~\cite{Dudek:2012vr} and 
from the future Electron-Ion-Collider to be built at 
Brookhaven National Lab~\cite{AbdulKhalek:2021gbh}. 
It is, thus, of great importance to extract TMDPDFs 
from first principles calculations, namely, lattice QCD.

In the last 10 years, there has been an enormous advance 
in computing the collinear PDFs using lattice QCD~\cite{Ji:2013dva,Radyushkin:2017cyf,Cichy:2018mum,Radyushkin:2019mye,Constantinou:2020pek,Ji:2020ect,Cichy:2021lih,Cichy:2021ewm,Constantinou:2022yye}.
By comparison, the computation of TMDPDFs is still in its infancy~\cite{Ji:2014hxa,Ji:2018hvs,Ji:2019sxk,Ebert:2022fmh}, 
although progress has been made in the computation of key 
elements required to build the TMDPDFS, like the soft functions~\cite{LatticeParton:2020uhz,Li:2021wvl} and the 
Collins-Soper kernel~\cite{Ebert:2018gzl,Ebert:2019tvc,Shanahan:2019zcq}.
Such computations  have been restricted to ratios of TMDPDFs~\cite{Hagler:2009mb,Musch:2011er,Yoon:2015ocs,Yoon:2017qzo}, 
and only recently, a first full lattice QCD calculation 
of the TMDPDFs themselves has been presented~\cite{LPC:2022zci}. 
One fundamental difficulty in  these calculations is to 
have control over the renormalization procedure, which 
is more involved than in the case of collinear PDFs. 
In particular, the staple-shaped gauge link that enters in the evaluation of TMDPDFs has three types of divergences: i) linear divergence coming from the Wilson line, which connects the quark fields and which depends on the length of the staple-shaped link; ii) logarithmic divergences coming form the endpoints of the staple link, similarly to the case of 
straight gauge links; and iii) logarithmic divergences coming from the presence of cusps in the staple. 
In addition, in the limit of infinite-length staple, $L$, (which is the case of interest), pinch-pole singularities arise as positive powers of $L$, coming from the gluon exchange between the transverse segments of the staple. 
Moreover, as in the case of straight gauge links, staple-shaped
operators of different Dirac structures can and will mix on the lattice among certain groups (when chiral symmetry is broken), as dictated by discrete symmetries.
However, in this case, the mixing pattern of the operators employed in lattice regularization can be significantly more 
involved than in the case of the straight Wilson line.

A first study within lattice perturbation theory to one-loop for the case of the symmetric staple~\cite{Constantinou:2019vyb} showed mixing between specific pairs of Dirac structures. This mixing cannot be avoided when one is interested in matching bare lattice Green's functions to the $\overline{\rm MS}$ scheme (directly or indirectly through an intermediate scheme). The mixing depended solely on the direction of the staple link entering the endpoints of the staple, regardless of the shape of the staple. This implies that the same mixing pattern occurs also for asymmetric staple shapes. In Ref.~\cite{Ji:2021uvr}, the mixing pattern of these staple-shaped operators\footnote{For conciseness, we will refer to staple-shaped Wilson-line quark bilinear operators as staple-shaped operators.} has been studied using symmetry considerations. It was found that more mixing is present than observed in Ref.~\cite{Constantinou:2019vyb}. This demonstrates that one-loop perturbation theory cannot fully reveal the mixing of the staple-shaped operators, unlike the case of straight Wilson-line operators~\cite{Constantinou:2017sej}, and a higher-loop computation is needed to confirm the additional mixing patterns found by symmetry arguments. The authors of Refs.~\cite{Shanahan:2019zcq,LatticeParton:2020uhz,Zhang:2022xuw} consider a maximal RI-type prescription, in which all 16 independent non-local Wilson-line quark bilinear operators are chosen to mix to eliminate all possible mixing effects. While Ref.~\cite{Shanahan:2019zcq} identified nonzero contributions in several off-diagonal elements of the renormalization matrix, Refs.~\cite{LatticeParton:2020uhz,Zhang:2022xuw} found negligible contributions, at least at small transverse separations, by setting specific momentum components to be zero. However, we emphasize that not all contributions are necessary for addressing the ``unavoidable'' mixing among the asymmetric staple-shaped operators on the lattice. Most off-diagonal elements in the $16 \times 16$ renormalization matrix are nonzero due to the non-minimal choice of renormalization conditions and not due to the unavoidable mixing. In this sense, it is preferable to construct a minimal intermediate scheme, keeping only the mixing sets that are needed for matching the bare lattice Green's functions to the corresponding $\overline{\rm MS}$-renormalized Green's functions (as obtained in dimensional regularization). In our study, we consider such a minimal scheme by using symmetry arguments to restrict the operators allowed to mix. \\
Improving renormalization schemes on the lattice eliminates finite lattice spacing errors, which can come from different Dirac structures in Green's functions under consideration. A way of removing artifacts from all Dirac structures is to consider a wider mixing pattern, where higher dimensional operators multiplied by the appropriate power of the lattice spacing can also mix with the operators under study. This mixing is only present for finite values of the lattice spacing, while it vanishes when taking the continuum limit. The higher-dimensional operators will be higher twist since (by Lorentz invariance) they must have the same spin (with the operators under study), but their dimension will be higher. The unwanted effects of finite lattice spacing errors and higher-twist contributions are not considered in the present study; they will be addressed in a future publication.

On the other hand, it has been shown~\cite{Zhang:2020rsx} 
that the linear divergence in the lattice spacing $a$ is not 
fully eliminated when the RI/MOM scheme is used to renormalize 
a straight Wilson line of length $z$. 
Ref.~\cite{Zhang:2022xuw} has shown that this residual 
linear divergence remains in the case of the staple-shaped 
operator. In this scenario, an 
alternative approach that one 
can adopt is the so-called {\it ratio scheme} as proposed for 
the quasi-PDF
case~\cite{Radyushkin:2017cyf,Orginos:2017kos,Braun:2018brg}.
In this approach, one subtracts the Ultra-Violet (UV) divergences
by taking the ratio with a suitable object at a fixed
short distance, where perturbation theory applies. 
Different choices of suitable objects
to be used in the ratio have been 
proposed in Ref.~\cite{Ji:2020brr}.
The authors of Refs.~\cite{Zhang:2022xuw,LPC:2022zci} use
ratios of the matrix elements of the operator under study.
As the divergences of the staple-shaped 
operator are independent of the longitudinal momentum of 
the state, one can use 
the matrix elements computed at different values of the momentum
in order to cancel the divergences. Usually, matrix elements at zero
momentum are chosen for the denominator, and such  scheme has been 
named~\cite{LPC:2022zci} {\it short distance ratio} (SDR) scheme.

As for the remaining 
divergences associated with the  asymmetric staple-shaped
link, one can cancel them by taking an appropriate 
ratio with the vacuum expectation value of a rectangular 
Wilson loop~\cite{Ji:2019ewn}.
However, we would like to stress that the SDR scheme is 
valid when operator mixing is absent or negligible. We 
check that in the case under study the mixing is indeed 
negligible and thus, one can employ the ratio scheme.
Another option, as proposed in Ref.~\cite{Ji:2021uvr}, 
is to employ RI/MOM in the spirit of the SDR scheme by 
fixing the dimensions of the staple at a short perturbative 
range (we will call this scheme RI-short). 

The paper is organized as follows.
Section~\ref{section:qTMD} discusses how TMDs can be accessed on a Euclidean lattice using the quasi-distribution approach.
In Section~\ref{section:operator_mixing}, 
we study  operator mixing using symmetry arguments. 
Section~\ref{section:lattice_setup} presents our lattice setup.
The size of the mixing is examined in Section~\ref{section:non_pert_renom}. 
In Sections~\ref{section:RI/MOM}, ~\ref{section:Ratio_scheme} 
and ~\ref{section:short_RI/MOM} we employ RI/MOM, SDR and 
RI-short schemes, respectively, by following the procedures described 
in Ref.~\cite{Zhang:2022xuw,LPC:2022zci,Ji:2021uvr}. 
In Section~\ref{section:results}, we present the renormalized beam 
functions in the three renormalization schemes discussed 
here, and then show the corresponding results in the $\MSbar$ scheme. 
The conclusions are presented in Section~\ref{section:conclusions}.

\section{Quasi-TMDPDFs}
\label{section:qTMD}

\begin{figure}[!h]
	\begin{center}
		\begin{tikzpicture}
			\draw[black, thick, ->] (0,0) -- (-2,0);
			\draw[black, thick] (-2,0) -- (-4,0);
			\draw[black, thick, ->] (-4,0) -- (-4,1);
			\draw[black, thick] (-4,1) -- (-4,2);
			\draw[black, thick, ->] (-4,2) -- (-2,2);
			\draw[black, thick] (-2,2) -- (2,2);
			\filldraw[black] (0,0) circle (0pt) node[anchor=west]{$\vec{0}$};
			\filldraw[black] (-4,0) circle (0pt) node[anchor=north]{$ - L \hat{z}$};
			\filldraw[black] (-4,2) circle (0pt) node[anchor=south]{$ - L \hat{z} + b \hat{y}$};
			\filldraw[black] (2,2) circle (0pt) node[anchor=west]{$ + b \hat{y} + z \hat{z}$};
            \draw[black,thick,->] (5,0) -- (5,0.5);
            \draw[black,thick,->] (5,0) -- (5.5,0);
            \filldraw[black] (5.5,0) circle (0pt) node[anchor=west]{$\hat{z}$};
            \filldraw[black] (5,0.5) circle (0pt) node[anchor=south]{$\hat{y}$};
		\end{tikzpicture}%
		\caption{The shape of the asymmetric staple defined in the operator of the quasi-beam function}
		\label{fig:staple}
	\end{center}
\end{figure}
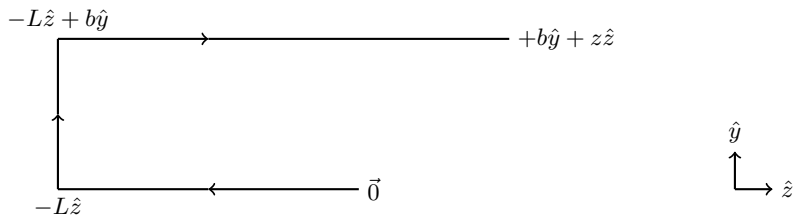

A first-principle computation of TMDPDFs in the context of 
LaMET is more involved than collinear PDFs. The main obstacle 
is to correctly subtract the so-called rapidity divergences, 
which are associated with the soft gluon radiation of the 
infinitely long Wilson lines present in the staple-shaped operator. 
This subtraction is made through the use of a soft function~\cite{Collins:2011ca,Echevarria:2015byo}. 
Different ways how to perform this subtraction in the context of 
LaMET can be found in Ref.~\cite{Ebert:2019okf}. In general, 
the TMD soft function involves two opposite light-like directions,
and this makes a lattice calculation significantly challenging. 
In Refs.~\cite{Ji:2019sxk,Ji:2019ewn}, the authors define a
rapidity-independent reduced soft function $S_r$ 
and show that it can be extracted from a form factor and 
the quasi-TMD wave function of a light meson. Using $S_r$, 
the rapidity scheme-independent TMDPDF $f^{TMD}(x,b,\mu,\zeta)$ 
can be written as~\cite{Ji:2019ewn,Ji:2020ect}
\begin{equation}
\begin{split}
	&f^{TMD}(x,b,\mu,\zeta) = H\left(\frac{\zeta_z}{\mu^2} \right) \, e^{-\ln{\left(\frac{\zeta_z}{\zeta}\right)} K(b,\mu)} \tilde{f}(x,b,\mu,\zeta_z)\, S_r^{\frac{1}{2}}(b,\mu) + \mathcal{O}\left(\frac{\Lambda_{QCD}^2}{\zeta_z},
    \frac{M^2}{(P^z)^2},\frac{1}{b^2\zeta_z} \right)
 \end{split}
\label{eq:TMDPDF}
\end{equation}
where $\tilde{f}(x,b,\mu,\zeta_z)$ is the so-called quasi-TMDPDF
for a nucleon with mass $M$, $x$ is the longitudinal momentum
fraction, and $b$ is a separation transverse to direction of 
the momentum $P^z$ carried by the nucleon.
The scale $\mu$ defines the renormalization scale and 
$\zeta_z = (2 x P^z)^2$ is the Collins-Soper scale of the 
quasi-TMDPDF, with $\zeta$ being the scale for the light-cone
correlation. The factor $H\left(\frac{\zeta_z}{\mu^2} \right)$ 
is the perturbative matching kernel that connects the TMDPDFs
to the quasi-TMDPDFs, and $K(b,\mu)$ is the Collins-Soper kernel. 
The renormalization scheme in Eq.~(\ref{eq:TMDPDF}) is left 
unspecified; usually, it is computed in the $\MSbar$ scheme.
The quasi-TMDPDF on the lattice is defined as~\cite{Ebert:2019okf,Ji:2018hvs}
\begin{equation}
	\tilde{f}(x,b,\mu,\zeta_z) =  \int \frac{d z}{2 \pi} e^{-i z \zeta_z} \frac{P^z}{E_P} B_{\Gamma} (z,b,\mu,P^z),
\end{equation}
with $B_{\Gamma} (z,b,\mu,P^z)$ the renormalized
beam function. It is obtained 
from the bare beam function, $B_{0,\Gamma}(z,b,L,P^z;1/a)$, defined
as the matrix element of the non-local staple-shaped 
quark bilinear operator, $\mathcal{O}^\Gamma (z,b,L)$,
of length $L$:
\begin{eqnarray}
		B_{0,\Gamma} (z,b,L,P^z;1/a) &=& \langle N(P^z) \vert \mathcal{O}^\Gamma (z,b,L) \vert N(P^z) \rangle \nonumber \\ 
		&=& \langle N(P^z) \vert \bar{\psi}(b\hat{y} + z\hat{z}) 
          \Gamma \mathcal{W}(z,b,L;1/a) \tau_3 \psi(0) \vert N(P^z) \rangle,
	\label{eq:B_def}
\end{eqnarray}
where $N(P^z)$ is a nucleon state with momentum boost of 
$(0, 0, P^z)$, $a$ is the lattice spacing, $\mathcal{W}(z,b,L;1/a)$ is the staple-shaped
Wilson line, and $\psi(z)$ is the standard up and down quark doublet.
In practice, we compute the flavor non-singlet combination
$u-d$, and hence the operator also includes a Pauli $\tau_3$ 
matrix in flavor space.
For the unpolarized TMDPDF, $\Gamma$ can be either $\gamma_0$ 
or $\gamma_3$ or, in general, a combination of the two.
In this work, we show results for $\Gamma = \gamma_0$,
since we observe a better signal for this operator compared 
to $\Gamma = \gamma_3$. $\gamma_0$ was also used for the 
quasi-PDF case,
in which case mixing is absent, while $\gamma_3$ mixed 
with $\mathbf{1}$. In the case of the staple, however, no
such advantage exists, as will be shown in 
Section~\ref{section:operator_mixing}.
In general, $\mathcal{W}(z,b,L;1/a)$ is given by:
\begin{equation}
	\mathcal{W}(z,b,L) = W_z (\vec{0},-L) W_\perp (-L\hat{z},b) 
        W_{z} (-L\hat{z} + b\hat{y}, L + z)\,
\label{eq:staple_shaped_GL}
\end{equation}
%
where we dropped the dependence on $1/a$ 
to keep the notation compact.
The arguments of the Wilson lines on the right 
hand side (RHS) of the equation above are defined 
in the following way: the indexes $z$ and $\perp$ in $W_z$ 
and $W_\perp$ refers to the direction $\hat{z}$ of the boost 
and a direction perpendicular to it, respectively. 
The transverse direction could be in the $\hat{x}$ or 
$\hat{y}$ direction. 
The first argument of both   $W_z$ and $W_\perp$ refers
to the initial point of the Wilson line, while the second 
argument is the displacement of the Wilson
line in the direction of the indexes ($z$ or $\perp$).
Mathematically, the expression for a Wilson line starting
at the point $\vec{x}$, with a displacement $L$ in the 
$\hat{z}$ direction, is given by:
\begin{equation}
    W_z(\vec{x};L) = {\cal P} \exp \left[-ig\int_0^L d\lambda \hat{z}\cdot A(\vec{x}+\hat{z}\lambda)\right],
\end{equation}
with $\cal P$ the path order operator and $A$ the gluon
field.
The shape of the asymmetric staple is shown in 
Fig.~\ref{fig:staple}.
%

\section{Operator mixing through symmetry}
\label{section:operator_mixing}

\begin{figure}[!h]
	\begin{center}
		\begin{tikzpicture}
            \draw[black, thick,
            decoration={markings,mark=at position 1 with {\arrow[red]{>}}},
    postaction={decorate},
    shorten >=0.4pt] (-2,0) -- (-2,2/2);
            \draw[black, thick, <-] (-2,2/2) -- (-2,4/2);
            \draw[black, thick,
            decoration={markings,mark=at position 1 with {\arrow[red]{>}}},
    postaction={decorate},
    shorten >=0.4pt] (-2,4/2) -- (-3/2,4/2);
            \draw[black, thick, <-] (-3/2,4/2) -- (-1,4/2);
            \draw[black, thick,
            decoration={markings,mark=at position 1 with {\arrow[red]{>}}},
    postaction={decorate},
    shorten >=0.4pt] (-1,4/2) -- (-1,3/2);
            \draw[black, thick, <-] (-1,3/2) -- (-1,2/2);
            \filldraw[black] (-2.3,-0.2) circle (0pt) node[anchor=west]{$q(0,0)$};
            \filldraw[red] (-2.3,-0.6) circle (0pt) node[anchor=west]{$\overline{q}(0,0)$};
            \filldraw[black] (-7/2,2.3) circle (0pt) node[anchor=west]{$(0,L+z)$};
            \filldraw[black] (-1.2,2.3) circle (0pt) node[anchor=west]{$(b,L+z)$};
            \filldraw[black] (-1.3,-0.2+1) circle (0pt) node[anchor=west]{$\overline{q}(b,z)$};          
            \filldraw[red] (-1.3,-0.6+1) circle (0pt) node[anchor=west]{$q(b,z)$};
            \filldraw[black] (-4,1.5) circle (0pt) node[anchor=west]{$O^{++}(\Gamma)$}; 
            \filldraw[red] (-4,1) circle (0pt) node[anchor=west]{$O^{++}_c(\Gamma)$};
            \draw[black, thick,
            decoration={markings,mark=at position 1 with {\arrow[red]{>}}},
    postaction={decorate},
    shorten >=0.4pt] (2,2) -- (2,1);
            \draw[black, thick, <-] (2,1) -- (2,0);
            \draw[black, thick,
            decoration={markings,mark=at position 1 with {\arrow[red]{>}}},
    postaction={decorate},
    shorten >=0.4pt] (2,0) -- (5/2,0);
            \draw[black, thick, <-] (5/2,0) -- (3,0);
            \draw[black, thick,
            decoration={markings,mark=at position 1 with {\arrow[red]{>}}},
    postaction={decorate},
    shorten >=0.4pt] (3,0) -- (3,1/2);
            \draw[black, thick, <-] (3,1/2) -- (3,1);
            \filldraw[black] (1.3,2.2) circle (0pt) node[anchor=west]{$q(0,0)$};
            \filldraw[red] (1.3,2.6) circle (0pt) node[anchor=west]{$\overline{q}(0,0)$};
            \filldraw[black] (0.3,-0.3) circle (0pt) node[anchor=west]{$(0,-L-z)$};
            \filldraw[black] (2.8,-0.3) circle (0pt) node[anchor=west]{$(b,-L-z)$};
            \filldraw[black] (2.5,1.2) circle (0pt) node[anchor=west]{$\overline{q}(b,-z)$};          
            \filldraw[red] (2.5,1.6) circle (0pt) node[anchor=west]{$q(b,-z)$};
            \filldraw[black] (4,1.5) circle (0pt) node[anchor=west]{$O^{+-}(\Gamma)$}; 
            \filldraw[red] (4,1) circle (0pt) node[anchor=west]{$O^{+-}_c(\Gamma)$};

		\end{tikzpicture}\\
		\begin{tikzpicture}
            \draw[black, thick, ->] (-2,2/2) -- (-2,3/2);
            \draw[black, thick,
            decoration={markings,mark=at position 0.2 with {\arrow[red]{<}}},
    postaction={decorate},
    shorten >=0.4pt] (-2,3/2) -- (-2,4/2);
            \draw[black, thick, ->] (-2,4/2) -- (-3/2,4/2);
            \draw[black, thick,
            decoration={markings,mark=at position 0.2 with {\arrow[red]{<}}},
    postaction={decorate},
    shorten >=0.4pt] (-3/2,4/2) -- (-1,4/2);
            \draw[black, thick, ->] (-1,4/2) -- (-1,2/2);
            \draw[black, thick,
            decoration={markings,mark=at position 0.2 with {\arrow[red]{<}}},
    postaction={decorate},
    shorten >=0.4pt] (-1,2/2) -- (-1,0);
            \filldraw[black] (-1.3,-0.2) circle (0pt) node[anchor=west]{$q(0,0)$};
            \filldraw[red] (-1.3,-0.6) circle (0pt) node[anchor=west]{$\overline{q}(0,0)$};
            \filldraw[black] (-7/2,2.3) circle (0pt) node[anchor=west]{$(0,L+z)$};
            \filldraw[black] (-1.2,2.3) circle (0pt) node[anchor=west]{$(-b,L+z)$};
            \filldraw[black] (-2.6,-0.25+1) circle (0pt) node[anchor=west]{$\overline{q}(-b,z)$};          
            \filldraw[red] (-2.6,-0.65+1) circle (0pt) node[anchor=west]{$q(-b,z)$};
            \filldraw[black] (-4,1.5) circle (0pt) node[anchor=west]{$O^{-+}(\Gamma)$}; 
            \filldraw[red] (-4,1) circle (0pt) node[anchor=west]{$O^{-+}_c(\Gamma)$};
            \draw[black, thick, ->] (2,1) -- (2,1/2);
            \draw[black, thick,
            decoration={markings,mark=at position 0.2 with {\arrow[red]{<}}},
    postaction={decorate},
    shorten >=0.4pt] (2,1/2) -- (2,0);
            \draw[black, thick, ->] (2,0) -- (5/2,0);
            \draw[black, thick,
            decoration={markings,mark=at position 0.2 with {\arrow[red]{<}}},
    postaction={decorate},
    shorten >=0.4pt] (5/2,0) -- (3,0);
            \draw[black, thick, ->] (3,0) -- (3,1);
            \draw[black, thick,
            decoration={markings,mark=at position 0.2 with {\arrow[red]{<}}},
    postaction={decorate},
    shorten >=0.4pt] (3,1) -- (3,2);
            \filldraw[black] (1,1.2) circle (0pt) node[anchor=west]{$\overline{q}(-b,-z)$};
            \filldraw[red] (1,1.6) circle (0pt) node[anchor=west]{$q(-b,-z)$};
            \filldraw[black] (0.3,-0.3) circle (0pt) node[anchor=west]{$(0,-L-z)$};
            \filldraw[black] (2.8,-0.3) circle (0pt) node[anchor=west]{$(-b,-L-z)$};
            \filldraw[black] (2.5,2.2) circle (0pt) node[anchor=west]{$q(0,0)$};          
            \filldraw[red] (2.5,2.6) circle (0pt) node[anchor=west]{$\overline{q}(0,0)$};
            \filldraw[black] (4,1.5) circle (0pt) node[anchor=west]{$O^{--}(\Gamma)$}; 
            \filldraw[red] (4,1) circle (0pt) node[anchor=west]{$O^{--}_c(\Gamma)$};

		\end{tikzpicture}
  \caption{All distinct possibilities for an asymmetric staple-shaped Wilson line
  connecting a quark and an antiquark field spatially separated by $\pm b\hat{y}$ and $\pm z\hat{z}$ ($b,z\geq0$).}
		\label{fig:symmetries}
	\end{center}
\end{figure}
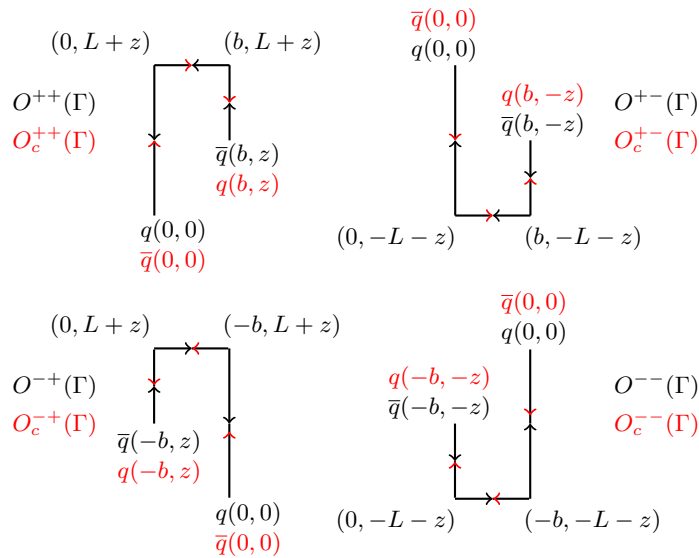

In order to use symmetry arguments to determine which operators mix, we
first organize the staple-shaped operator using all possible ways to
connect the fermion fields. We show in Fig.~\ref{fig:symmetries} 
all the 
distinct possibilities of connecting a quark and an antiquark field located at
a spatial separation of $(0,b,z)$, using an asymmetric
staple-shaped Wilson line. Since we build the staples only in the
$yz$-plane, in the following discussion and in Fig.~\ref{fig:symmetries}, we
denote the coordinate space $(t,x,y,z)$ as only $(y,z)$. Without loss of 
generality, we can fix the quark field ($q$) at $(0,0)$. Then, the only 
possibilities for the position of the antiquark field ($\bar{q}$) are $(b,z)$,
$(b,-z)$, $(-b,z)$ and $(-b,-z)$. Here, $b$ and $z$ are strictly non-negative.
We define the staple-shaped Wilson line connecting $q$ at $(0,0)$ and $\bar{q}$
at $(b,z)$ as $O^{++}(\Gamma)$. The two plus signs denote the direction of
separation of $\bar{q}$ from $q$. The first sign is for the transverse ($\hat{y}$)
and the second sign for the longitudinal ($\hat{z}$) axes. The $\Gamma$ denotes
the insertion gamma matrix as defined in Eq.~\ref{eq:B_def}. In a similar
fashion, the staple-shaped operator in the cases of $\bar{q}$ being located at
$(b,-z)$, $(-b,z)$ and $(-b,-z)$ can be defined as $O^{+-}(\Gamma)$, 
$O^{-+}(\Gamma)$ and $O^{--}(\Gamma)$ respectively. A visual representation
of these operators is shown in Fig.~\ref{fig:symmetries} in black. Due to
the asymmetric nature of the staple-shaped Wilson line, there are four more
operators that are obtained from the charge conjugation of the above defined four.
These are also shown in Fig.~\ref{fig:symmetries}, but in red, 
with the charge-conjugated version of $O^{\pm\pm}(\Gamma)$ denoted by $O^{\pm\pm}_c(\Gamma)$ and quark/antiquark fields having exchanged positions. 
For the symmetric
case ($z = 0$), the charge-conjugated operators are redundant, since
$O^{++}_c \equiv O^{-+}$, $O^{+-}_c \equiv O^{--}$, $O^{-+}_c \equiv O^{++}$
and $O^{--}_c \equiv O^{+-}$. 

We analyze the symmetry properties using generalized parity 
($\mathcal{P}_{F \alpha}^{1,2}$) and time reversal ($\mathcal{T}_{F \alpha}^{1,2}$)
transformations with discrete flavor rotation, as well as charge 
conjugation ($\mathcal{C}$), for the fermion fields in the twisted mass basis, 
$\chi(x_\alpha,\textbf{x})$, and with the gauge link, $U(x_\alpha,\textbf{x};\alpha)$,
in some direction $\alpha$. The symmetry transformations are defined in Appendix A. 
The operators $O^{\pm\pm}(\Gamma)$, $O^{\pm\pm}_c(\Gamma)$ do not have definite properties with respect to the symmetries of the lattice action. Instead, one needs to consider their linear combinations, defined by 
\begin{equation}
    (ijkl)_c = i\cdot O^{++} + j\cdot O^{--} + k\cdot O^{+-} + l\cdot O^{-+}
    + c \cdot \left(i\cdot O^{++}_c + j\cdot O^{--}_c + k\cdot O^{+-}_c + l\cdot O^{-+}_c\right),
    \label{combinations}
\end{equation}
with $i,j,k,l=\pm1$ denoting the signs of $O^{++}, O^{--}, O^{+-}, O^{-+}$, and $c = \pm1$ representing the relative sign of the charge-conjugated versions of $O^{\pm\pm}$.

The only combinations that have definite symmetry properties are:
\begin{align*}
    (++++)_c &\equiv (----)_c,\lbr
    (+-+-)_c &\equiv (-+-+)_c,\lbr
    (++--)_c &\equiv (--++)_c,\lbr
    (+--+)_c &\equiv (-++-)_c,
\end{align*}
i.e.\ the combinations with all signs $i,j,k,l$ reversed are equivalent from the point of view of symmetry transformations, with irrelevant global phase.

As an example, we look at the symmetry properties of $\gamma_0$ and
$\gamma_0 \gamma_3$ given in Tables~\ref{tab:sym_g0} and~\ref{tab:sym_g0g3},
respectively, restricting ourselves here to the 
flavor non-singlet case, $u-d$ ($\tau_3$ matrix in flavor space). The combinations of operators that mix are those which have all signs equal in the nine rows of Tables~\ref{tab:sym_g0} and~\ref{tab:sym_g0g3}. For example, the symmetry properties of $(++++)_c$ for $\Gamma=\gamma_0$ (second column of Table~\ref{tab:sym_g0}) are identical to the ones of the combination $(+--+)_c$ for $\Gamma=\gamma_0\gamma_3$ (last column of Table~\ref{tab:sym_g0g3}).
Thus, we conclude that the following mixings occur:
\begin{align*} 
    (++++)_c\;\; &\text{with}\;\; (+--+)_c,\lbr
    (+-+-)_c\;\; &\text{with}\;\; (++--)_c,\lbr
    (++--)_c\;\; &\text{with}\;\; (+-+-)_c,\lbr
    (+--+)_c\;\; &\text{with}\;\; (++++)_c,
\end{align*}
where the first combination in a pair pertains to $\Gamma=\gamma_0$ and the second one to $\Gamma=\gamma_0\gamma_3$. Additional mixings appear with $\Gamma=\gamma_0\gamma_2$ and $\Gamma=\gamma_5\gamma_1$, thus forming a quadruple of operators that mix, 
$\left\{\gamma_0,\gamma_0\gamma_2,\gamma_0\gamma_3,\gamma_5\gamma_1\right\}$.
These are the only relevant mixings for the present work, which concerns $\Gamma=\gamma_0$ for unpolarized TMDs.

\begin{table}[h]
\begin{tabular}{c c c c c}
    \hline
     &  \makecell{$(+ + + +)_c$\\$(- - - -)_c$} & 
     \makecell{$(+ - + -)_c$\\$(- + - +)_c$} & \makecell{$(+ + - -)_c$\\$(- - + +)_c$}
     & \makecell{$(+ - - +)_c$\\$(- + + -)_c$} \\[0.3cm]
     \hline
     $\mathcal{P}_{F0}^{1,2}$ & - & + & - & + \\[0.2cm]
     $\mathcal{P}_{F1}^{1,2}$ & + & - & + & - \\[0.2cm]
     $\mathcal{P}_{F2}^{1,2}$ & + & + & - & - \\[0.2cm]
     $\mathcal{P}_{F3}^{1,2}$ & + & - & - & + \\[0.2cm]
     \hline
     $\mathcal{T}_{F0}^{1,2}$ & + & + & + & + \\[0.2cm]
     $\mathcal{T}_{F1}^{1,2}$ & - & - & - & - \\[0.2cm]
     $\mathcal{T}_{F2}^{1,2}$ & - & + & + & - \\[0.2cm]
     $\mathcal{T}_{F3}^{1,2}$ & - & - & + & + \\[0.2cm]
     \hline
     $C$ & $c$ & $c$ & $c$ & $c$\\
     \hline
\end{tabular}
\caption{Symmetry properties of operators with $\Gamma=\gamma_0$. The $\pm$ sign for $\mathcal{P}_F/\mathcal{T}_F$ transformations denotes that a given combination of staple-shaped operators is symmetric/antisymmetric with respect to the symmetry transformation given in the first column. The last row indicates symmetry properties with respect to charge conjugation, which depend on the sign $c$, i.e.\ $(\cdot\cdot\cdot\cdot)_+$ combinations are symmetric and $(\cdot\cdot\cdot\cdot)_-$ antisymmetric with respect to $C$ in this case.}
\label{tab:sym_g0}
\vspace*{3mm}
\begin{tabular}{c c c c c}
    \hline
     & \makecell{$(+ + + +)_c$\\$(- - - -)_c$} & 
     \makecell{$(+ - + -)_c$\\$(- + - +)_c$} & \makecell{$(+ + - -)_c$\\$(- - + +)_c$}
     & \makecell{$(+ - - +)_c$\\$(- + + -)_c$} \\[0.3cm]
     \hline
     $\mathcal{P}_{F0}^{1,2}$ &  + & - & + & - \\[0.2cm]
     $\mathcal{P}_{F1}^{1,2}$ &  - & + & - & + \\[0.2cm]
     $\mathcal{P}_{F2}^{1,2}$ &  - & - & + & + \\[0.2cm]
     $\mathcal{P}_{F3}^{1,2}$ &  + & - & - & + \\[0.2cm]
     \hline
     $\mathcal{T}_{F0}^{1,2}$ & + & + & + & + \\[0.2cm]
     $\mathcal{T}_{F1}^{1,2}$ & - & - & - & - \\[0.2cm]
     $\mathcal{T}_{F2}^{1,2}$ & - & + & + & - \\[0.2cm]
     $\mathcal{T}_{F3}^{1,2}$ & + & + & - & - \\[0.2cm]
     \hline
     $C$ & $c$ & $c$ & $c$ & $c$\\
     \hline
\end{tabular}
\caption{Symmetry properties of operators with $\Gamma=\gamma_0\gamma_3$. See the caption of Fig.\ \ref{tab:sym_g0} for explanation.}
\label{tab:sym_g0g3}
\end{table}

For the general case, the symmetry properties for all $\Gamma$'s are summarized in \tab{tab:general_symmetry}
in Appendix A. They imply that the possible mixing is between $\Gamma$ and 
$\left\{\Gamma\gamma_2,\Gamma\gamma_3,\Gamma\gamma_2\gamma_3 \right\}$ in the asymmetric staples case.
In turn, symmetric staples restrict the mixing by eliminating always one member of the mixing quadruple, i.e.\ $\Gamma=\gamma_{0,1,5}$ mix with $\Gamma\gamma_2$ and $\Gamma\gamma_3$, while $\Gamma=\gamma_{2,3}$ form a triple $\left\{\gamma_2,\gamma_3, \gamma_2\gamma_3 \right\}$ of operators that mix.

We point out that different mixing patterns have been considered
in recent studies of other groups, including mixing among all 16 
operators of different Dirac structures~\cite{Shanahan:2019zcq}, 
or at least mixing in pairs ($\Gamma, \Gamma \gamma_3$)~\cite{Ji:2021uvr}. 
In our work, we choose to consider the minimal set of staple-shaped 
operators (of the same dimension) that are allowed to mix by 
the above-mentioned C, P, and T symmetries. We have not considered 
mixing with higher-dimensional operators allowed by Lorentz 
symmetry (see, e.g., Ref.~\cite{Chen:2017mie} for the straight Wilson-line 
case), since it is power suppressed and not relevant when one takes 
the continuum limit $a \rightarrow 0$. Also, in contrast to Ref.~\cite{Ji:2021uvr}, in 
our analysis, the staple line has been chosen to be in a 2-dimensional (D), and not 3D, 
plane in Euclidean space formed by the transverse ($\hat{y}$) and 
longitudinal ($\hat{z}$) directions. In this respect, we end up 
with a basis of 8 (instead of 16) operators, which are eigenstates 
of C, P and T transformations. 
We also note that calculations in one-loop lattice perturbation 
theory~\cite{Constantinou:2019vyb} show a smaller mixing pattern 
compared to the symmetries; however, this cannot guarantee a 
reduced mixing in higher loops.

\section{Lattice Setup}
\label{section:lattice_setup}

For the lattice simulation, we use an $N_f = 2 + 1 + 1$ 
clover-improved twisted mass fermion ensemble of size 
$24^3 \times 48$, produced by the Extended Twisted Mass 
Collaboration (ETMC)~\cite{Alexandrou:2018egz}.
Our study is done using $N_{conf}$ configurations with 
$N_{src}$ source positions for each configuration. To 
increase statistics, the boost is taken in all three  
directions, and both positive and negative. For each such 
direction of boost, the staple is then constructed in both 
the remaining transverse directions. This gives us 12 
measurements ($6$ boost directions $\times$ $2$ 
transverse directions) for each source position. The 
details of the lattice simulation are summarized 
in~\tab{tab:lattice_setup}.

\begin{table}[h!]
	\begin{tabular}{|c|c|c|c|c|c|c|}
		\hline
		Lattice size & $a$ [fm] & $a \mu_l$ & $m_\pi$ [MeV] & $N_{conf}$ & $N_{src}$ & $N_{meas}$\\
		\hline \hline
		$24^3 \times 48$ & $0.093$ & $0.00530$ & $350$ & $100$ & 8 & 9600\\
		\hline
	\end{tabular}
	\caption{We give the parameters of the  lattice ensemble and measurements used in the calculation. $a$ is the lattice spacing, $\mu_l$ is the bare twisted light quark mass,  $m_\pi$ the pion mass, $N_{conf}$ the number of configurations, $N_{src}$ the number of source positions and $N_{meas}$ the total number of measurements.}
	\label{tab:lattice_setup}
\end{table}

\begin{figure}[h]
	\centering
	\subfigure{\includegraphics[width=0.49\textwidth]{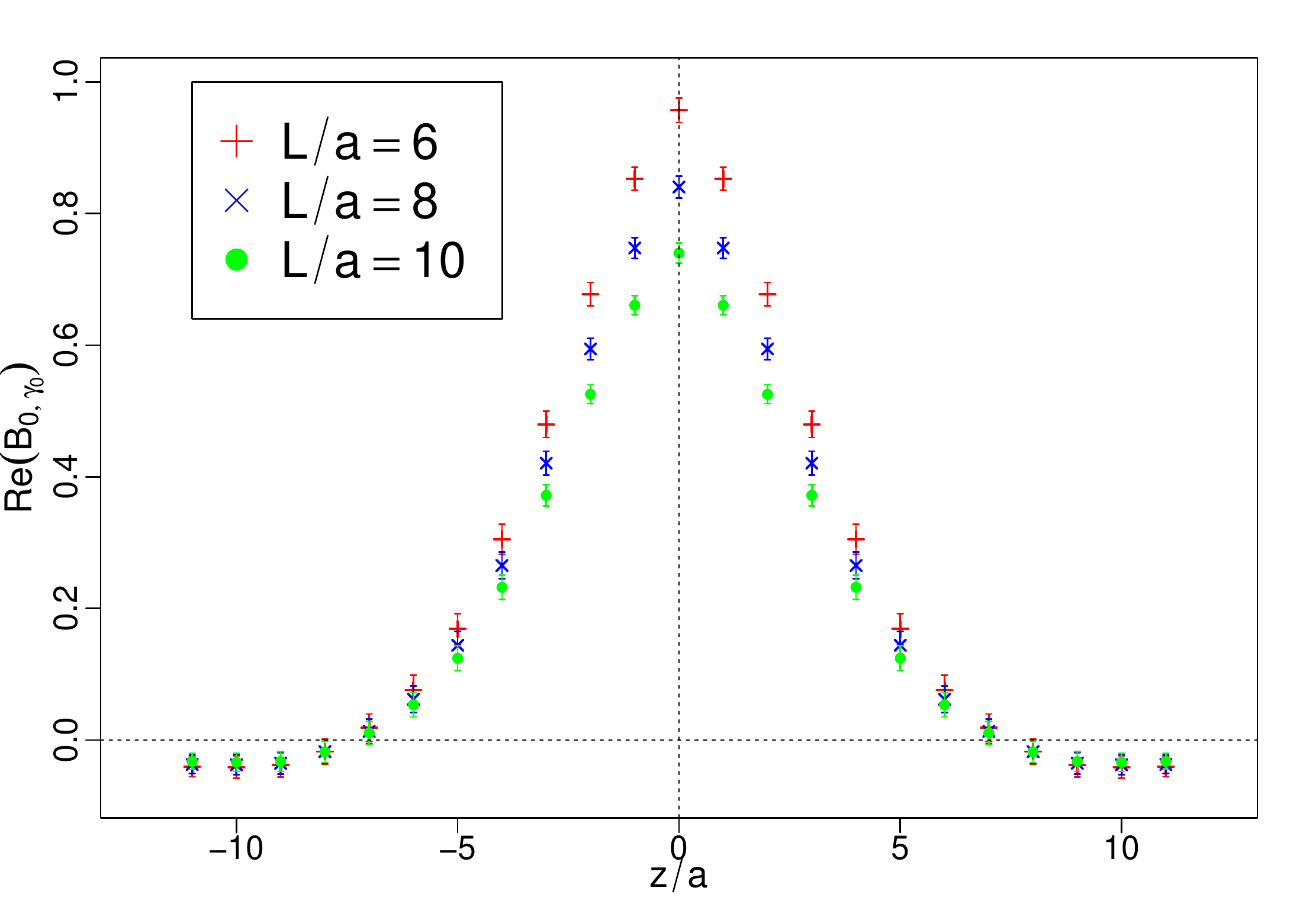}}
	\subfigure{\includegraphics[width=0.49\textwidth]{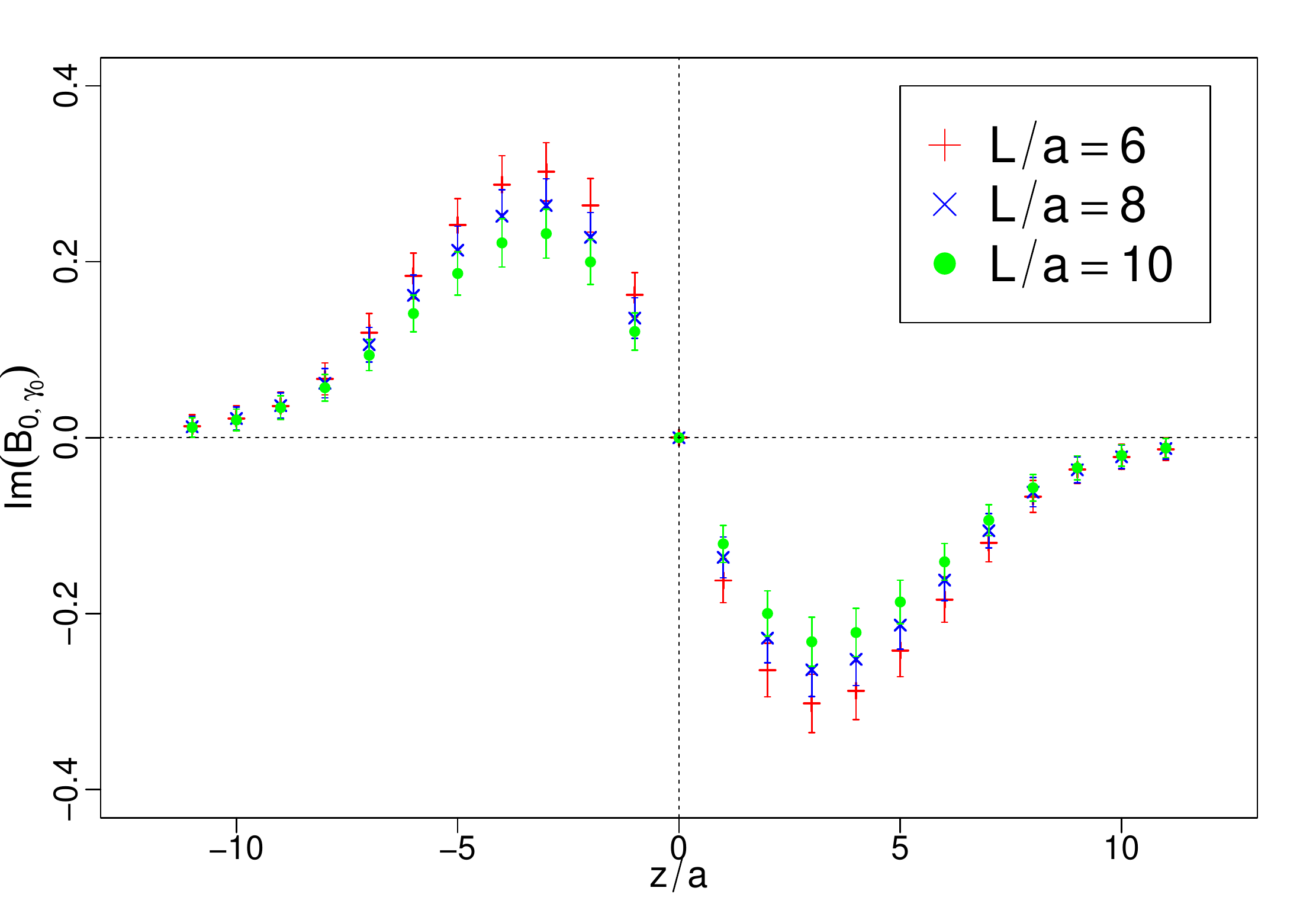}}
	\caption{Real and imaginary parts of the bare matrix elements for a transverse separation of $b/a = 1$.}
	\label{fig:B_bare}
\end{figure}

\begin{figure}[h]
    \centering
    \includegraphics[width=0.6\textwidth]{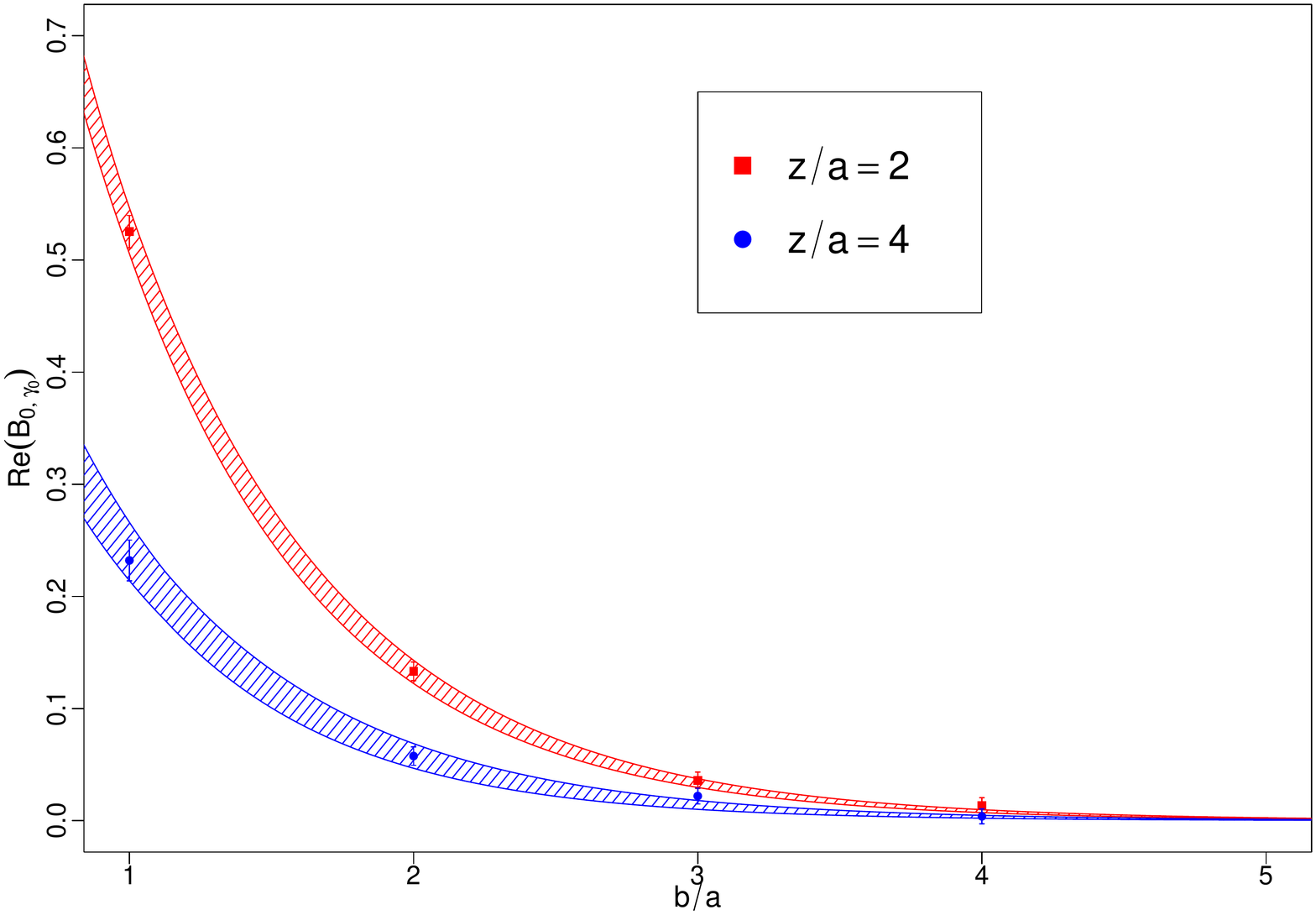}
    \caption{Exponential decay of the bare matrix element with increasing
    $b$ at $L = 10a$ and z/a = 2 (red points) and z/a = 4 (blue points). 
    The red and blue bands are the fits to the results at z/a = 2 and z/a = 4, 
    respectively.}
    \label{fig:B_vs_b}
\end{figure}

The bare matrix element for the quasi-beam function is calculated through a ratio of a 3-point to a 2-point function, 
\begin{equation}
	\begin{split}
		B_0^\Gamma (z,b,L,P^z) &= \frac{\langle C^{3pt}_\Gamma (z,b,L,P^z;t_s,\tau) \rangle}{\langle C^{2pt} (P^z;t_s) \rangle}\\
		&= \frac{{\cal P}\sum_\mathbf{x} e^{-i \mathbf{P} \cdot \mathbf{x}} \langle 0 \vert N(\mathbf{x},t_s) \mathcal{O}^\Gamma(z,b,L;\tau) \bar{N}(\mathbf{0},0) \vert 0 \rangle}{{\cal P}\sum_\mathbf{x} e^{-i \mathbf{P} \cdot \mathbf{x}} \langle 0 \vert N(\mathbf{x},t_s) \bar{N}(\mathbf{0},0) \vert 0 \rangle},
	\end{split}
\end{equation}
where $\tau$ is the insertion time of the operator $\mathcal{O}^\Gamma$ and $t_s$ defines 
the source-sink time separation. ${\cal P}$ is the parity projector. 
In this work, we show results for a single source-sink separation of $t_s = 10 a$ and a
longitudinal momentum of $P^z = 6\pi/24a \sim 1.7 \gev$.\\
The 3-point function is constructed for the isovector combination $u-d$, by inserting a $\tau_3$ in 
flavor space. This choice ensures the elimination of the disconnected contributions 
and only connected diagrams need to be calculated.\\
Momentum smearing~\cite{Bali:2016lva} is applied to improve 
the signal for large boosts. It is observed that applying stout 
smearing to the gauge links used in construction of the staple also reduces the statistical errors. 
Here, we have applied 5 steps of stout smearing to the staple-shaped Wilson line.\\
In Fig.~\ref{fig:B_bare} we show the bare matrix 
elements as a function of $z$ at a fixed $b/a=1$ for different values
of $L$. 
These and the following results for the quasi-beam function have been
symmetrized using the relation
\begin{equation}
    B_{0,\Gamma}(z,b,L,P^z) = B_{0,\Gamma}^\dagger(-z,-b,-L,P^z).
\end{equation}
To ensure that this property holds in our lattice simulation, we have used 
the staple-shaped operator $O^{++}$ on the left hand side of the above equation
and $O^{--}_c$ on the right hand side. 
Our results show a clear decay of the magnitude of the beam function 
as $L$ increases, as expected~\cite{Ji:2020ect}.
In Fig.~\ref{fig:B_vs_b} we show the dependence of the bare beam function
on $b$. This is expected to decay as $\exp{(-LV(b))}$~\cite{Ji:2020ect}.
Assuming a linear function for the potential $V(b)$, we also show a fit of $\exp{(c_0 + c_1 b})$
to the lattice data at two different values of $z/a$ and a fixed $L/a = 10$.
%

\section{Non-perturbative renormalization}
\label{section:non_pert_renom}

As in the case of the straight Wilson line, there are logarithmic and linear
divergences associated with the length $z$ of the staple-shaped 
link. However, here there are two extra divergences: one,
which is associated with the cusps of the staple, and one associated
with the gluon exchange between the gauge links of length $L\rightarrow\infty$,
the so-called pinch-pole singularities. On the other hand, the finite transverse
separation $b$ mitigates possible discretization effects associated
with small $z$ separation, mainly in the non-perturbative region (large $b$),
which is the main interest of a lattice calculation. In the following,
we will analyze three different ways to carry out the renormalization, 
namely the RI/MOM scheme, the short distance ratio scheme, and a modified version
of the RI/MOM scheme, where the renormalization constants are computed
at short distances.

\subsection{RI/MOM}
\label{section:RI/MOM}

\begin{figure}[!h]
    \centering
    \subfigure{\includegraphics[width=0.49\textwidth]{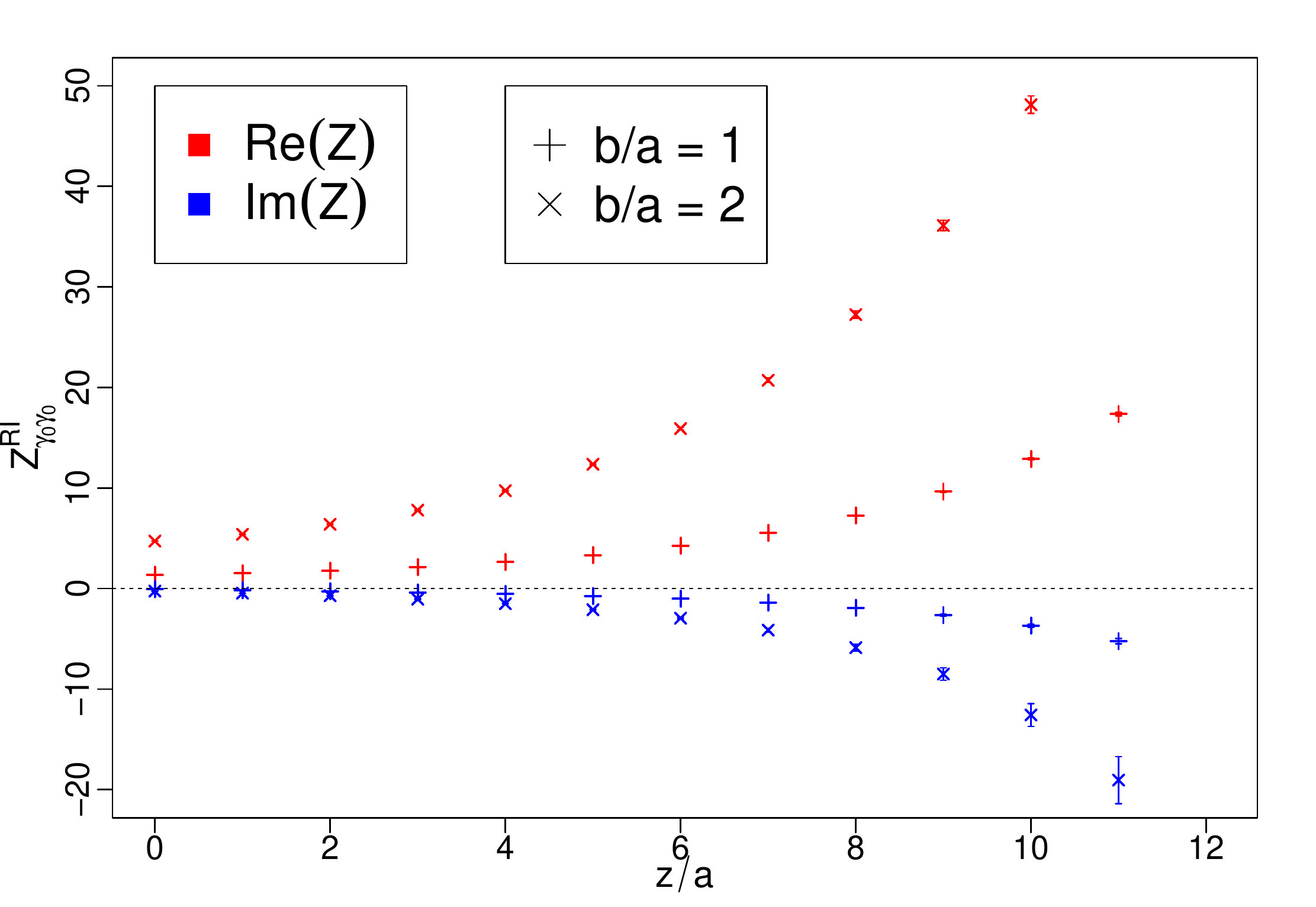}}
    \caption{RI/MOM renormalization factor as a function of the Wilson line length $z$ for 
    $b/a=1,2$ and at fixed $L/a = 10$. Both real and imaginary parts
    increase rapidly as $z$ grows.}
    \label{fig:Z_RI_g0}
\end{figure}

\begin{figure}[!h]
    \centering
    \subfigure{\includegraphics[width=0.49\textwidth]{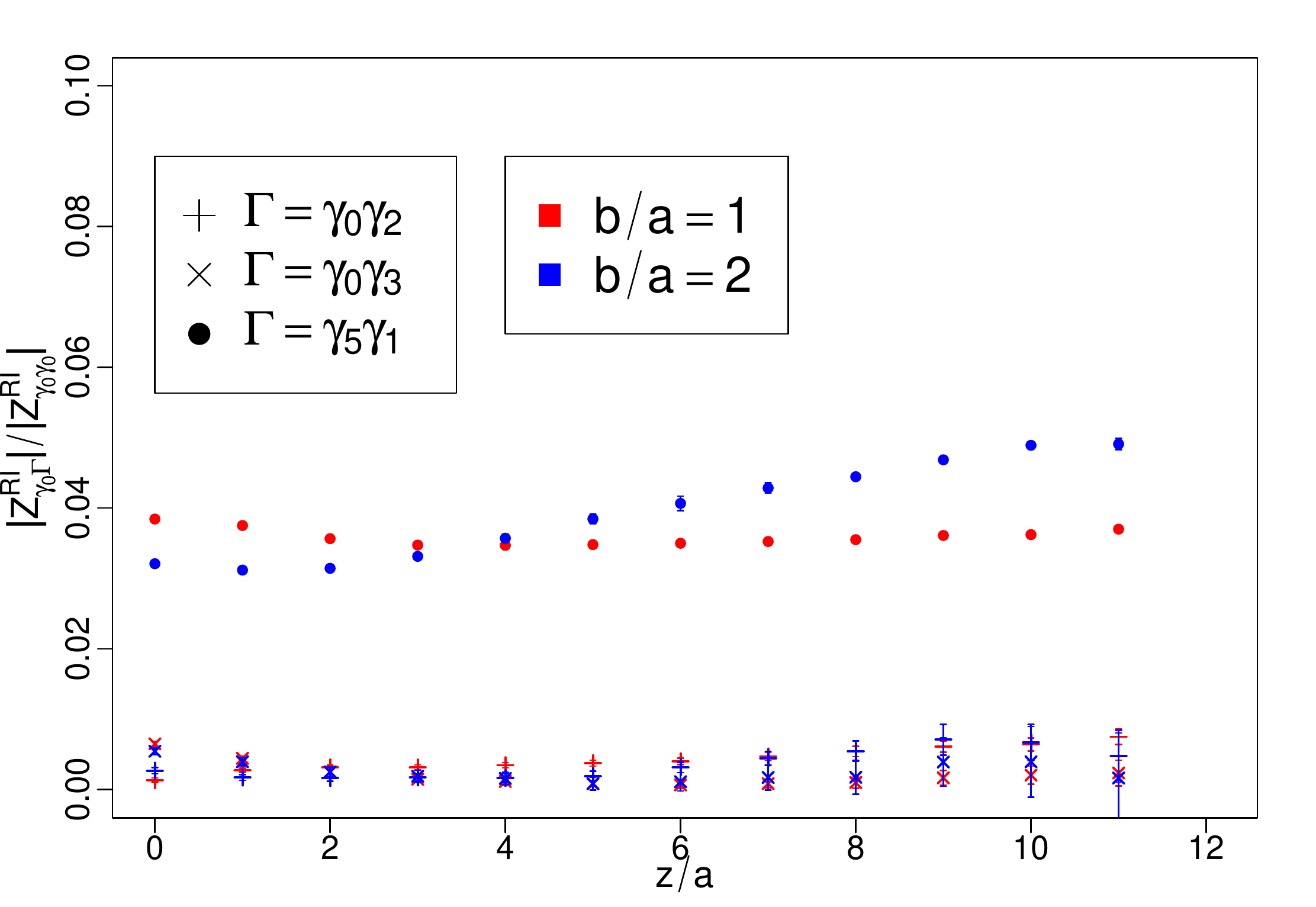}}
    \caption{Contribution of the off-diagonal mixing terms compared to the diagonal one, in the RI/MOM renormalization matrix at fixed $L/a = 10$.}
    \label{fig:Z_RI_g0_off}
\end{figure}

The RI/MOM scheme~\cite{Martinelli:1994ty} was first adapted for non-local operators employed in the quasi-distribution approach in Refs.~\cite{Constantinou:2017sej,Alexandrou:2017huk}.
The RI/MOM renormalization constants $Z^{RI}_{\cal O}$ 
are defined by the condition
\begin{equation}
    \frac{Z^{\RI}_{\Gamma\Gamma^{'}}(z,b,L,\mu_0;1/a)}{Z^{\RI}_q(\mu_0;1/a)}\frac{1}{12} {\rm Tr} \left[\frac{\Lambda^\Gamma_0(z,b,L,p;1/a)\Gamma^{'}}{e^{ip^zz+i p_\perp b}}\right] \Bigg|_{p{=}\mu_0} {=} 1\, ,
\end{equation} 
%
where $\Lambda^\Gamma_0$ is defined in terms of the amputated Green's function
\begin{equation}
    \Lambda^\Gamma_0(z,b,L,p;1/a) = S_q^{-1}G^\Gamma(z,b,L,p;1/a) S_q^{-1}\, ,
\end{equation}
with $S_q$  the off-shell quark propagator. The 
Green's function is calculated as
\begin{equation}
    G^\Gamma(z,b,p,L;1/a)=\langle q(p)| {\cal O}^\Gamma (z,b,p,L;1/a) |q(p)\rangle\, .
\end{equation}
Because $G^\Gamma$, and thus $\Lambda_0^\Gamma$, have the 
same divergences as $B_{0,\Gamma}$, all the divergences, in 
principle, cancel in the renormalization 
procedure.  $Z^{\RI}_q$ is the quark wave function 
renormalization defined as
\begin{equation}
    Z^{\RI}_q(\mu_0;1/a)= \frac{1}{12} {\rm Tr} 
    \left[(S_q(p;1/a))^{-1}\, S_q^{\rm Born}(p)\right] \Bigg|_{p^2=\mu_0^2}  \,,
\end{equation}
and the corresponding renormalized beam function is then 
\begin{equation}
    B^{\RI}_\Gamma (z,b,\mu_0,P^z)=
    \sum_{\Gamma^{'}}Z^{\RI}_{\Gamma\Gamma^{'}}(z,b,\mu_0;1/a)B_{0,\Gamma^{'}}(z,b,P^z;1/a)\,.
\end{equation}
Notice that the relation between the bare and renormalized 
beam functions involves, in principle, a $16\times 16$ matrix,
if one considers all possible mixing among the full set of Dirac 
structures.
The case of interest in this work is   $\Gamma=\gamma_0$,
computed using the lattice ensemble of Table~\ref{tab:lattice_setup}. 
As pointed out in Section \ref{section:operator_mixing},
$\gamma^0$ mixes with $\gamma_0\gamma_2$, $\gamma_0\gamma_3$, 
and $\gamma_5\gamma_1$. The renormalization factors for the 
diagonal and off-diagonal cases are shown in Figs.~\ref{fig:Z_RI_g0} 
and~\ref{fig:Z_RI_g0_off}, respectively. 
We set the renormalization scale 4-vector $\mu_0$ equal to 
$2 \pi \left( \frac{6 + 0.5}{N_t}, \frac{3}{N_s}, \frac{3}{N_s}, \frac{3}{N_s}\right)$, 
where $N_s = 24 a$, $N_t = 2 N_s$, and anti-periodic boundary 
conditions have been employed in the time direction. 
We have chosen an isotropic momentum in the spatial directions 
and ``democratic'' momentum, which obeys the criterion 
$\sum_\rho \sin^4 (a p_\rho/2) / (\sum_\rho \sin^2(a p_\rho/2))^2 |_{p = \mu_0} < 0.3$,
in order to reduce Lorentz-non-invariant contributions 
in the vertex functions. Also, the selected momentum lies in 
a perturbative region, where perturbation theory is reliable 
and, at the same time, lattice artifacts are under control. 
A more detailed study using different values of $\mu_0$ will 
be presented in a future extension of our study.
Using this set-up, we observe that
the contributions from the off-diagonal mixing terms are $\lesssim 4\%$ 
of the diagonal contribution for all $z$ when $b = 1a,\,2a$.

\begin{figure}[!h]
    \centering
    \subfigure{\includegraphics[width=0.49\textwidth]{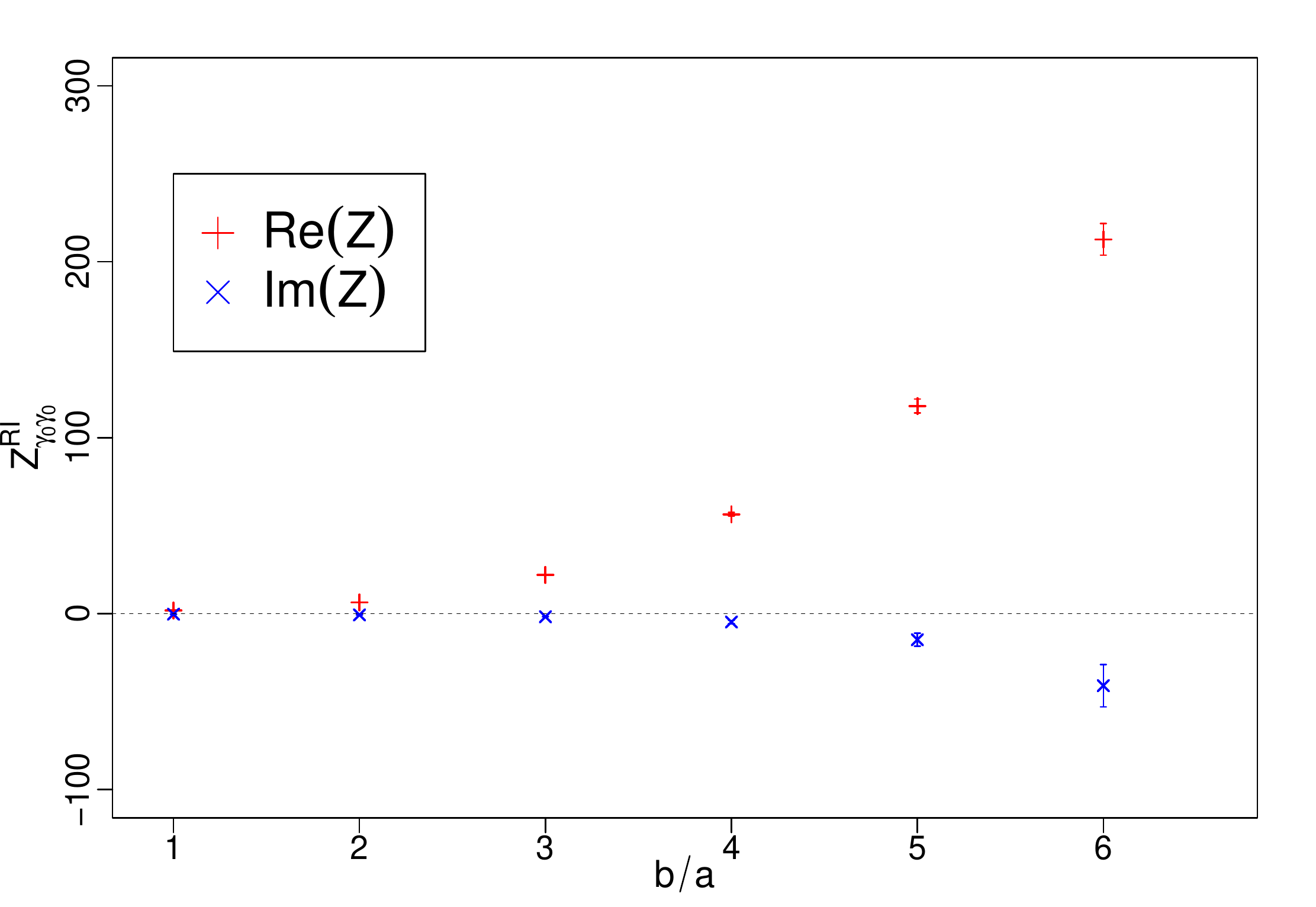}}
    \subfigure{\includegraphics[width=0.49\textwidth]{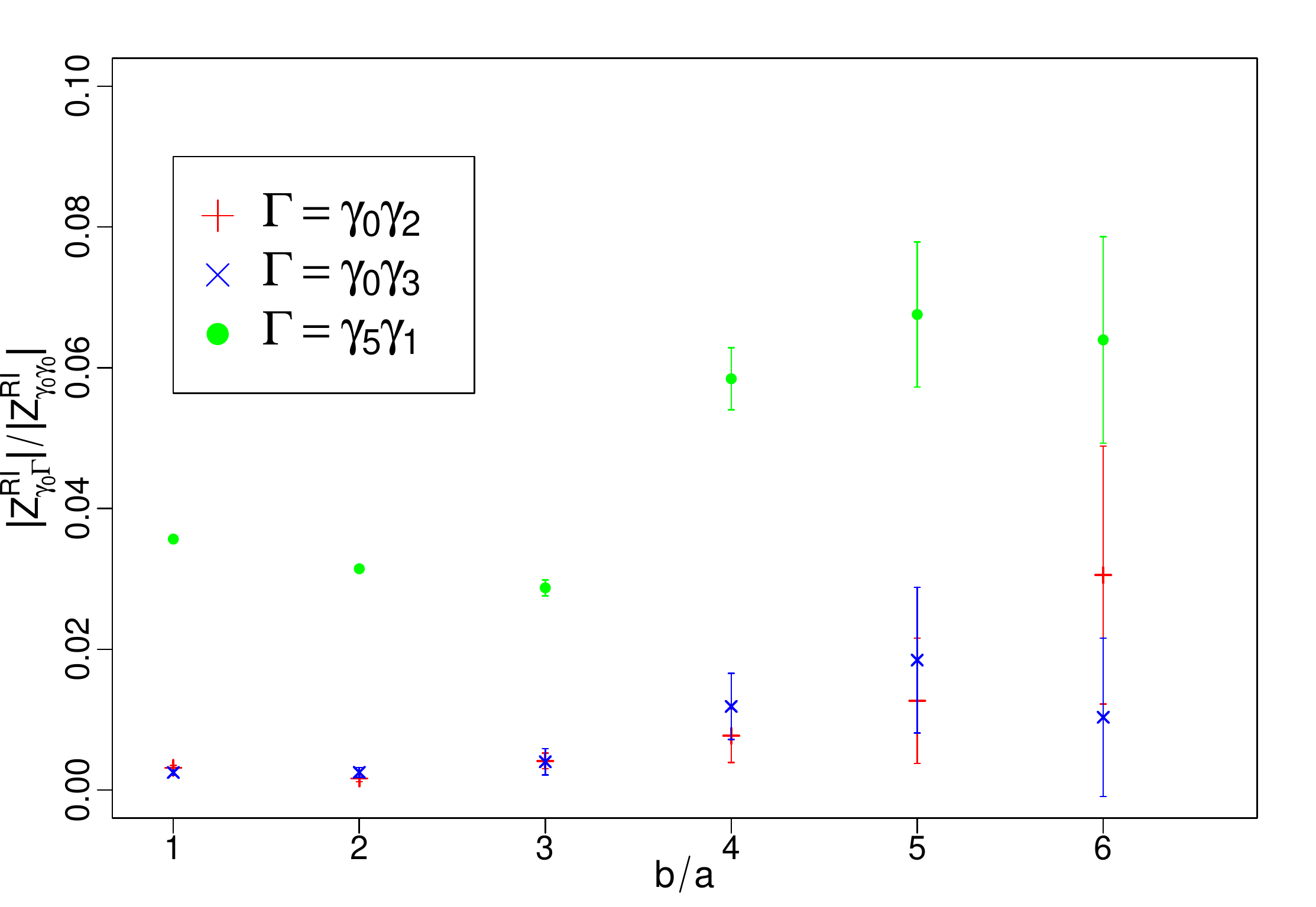}}
    \caption{The diagonal and off-diagonal contributions to the RI/MOM renormalization factors as a function of $b$ at fixed $L/a = 10$ and $z/a = 2$.}
    \label{fig:Z_RI_g0_large_b}
\end{figure}

\begin{figure}[h]
	\centering
	\subfigure{\includegraphics[width=0.49\textwidth]{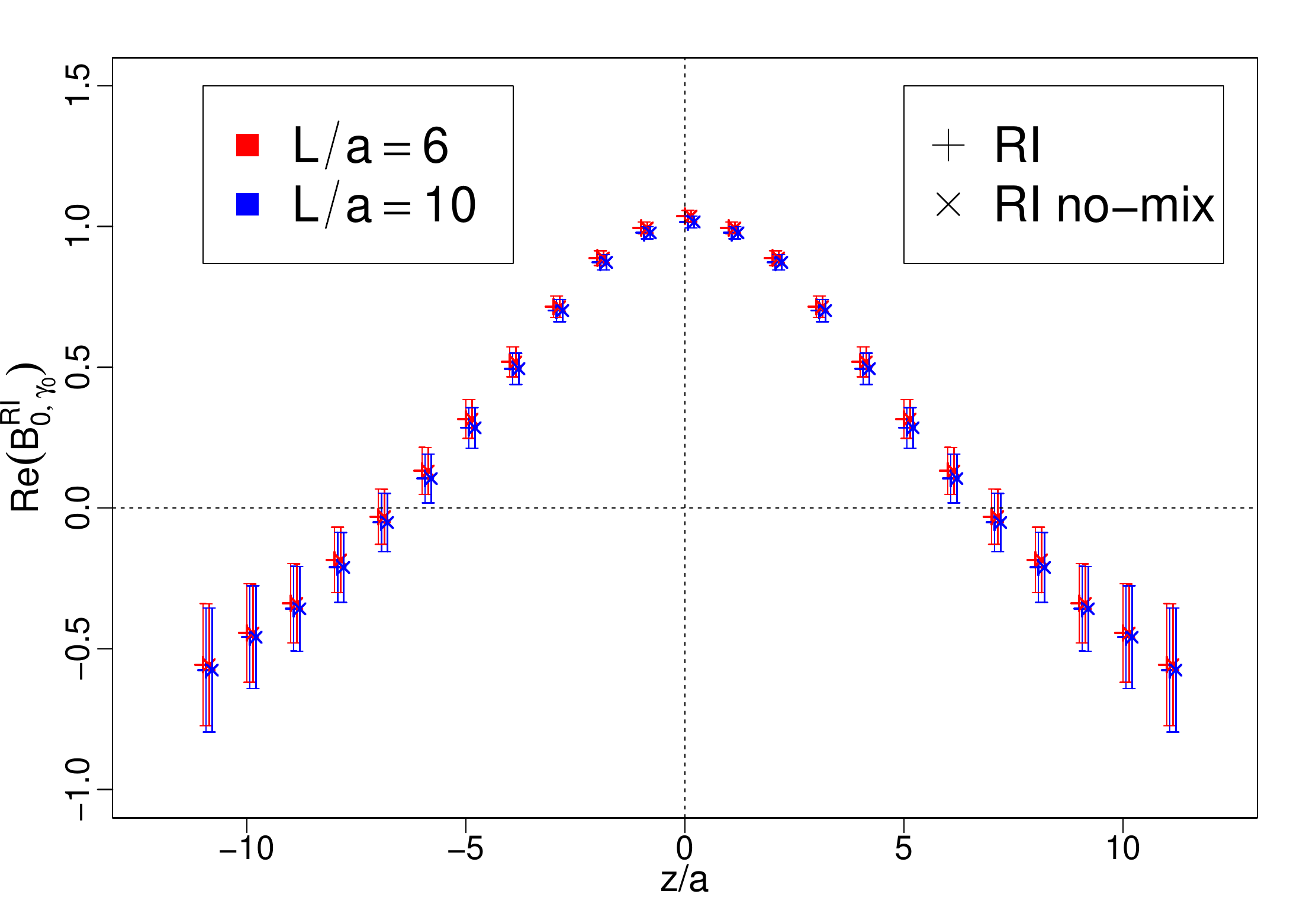}}
	\subfigure{\includegraphics[width=0.49\textwidth]{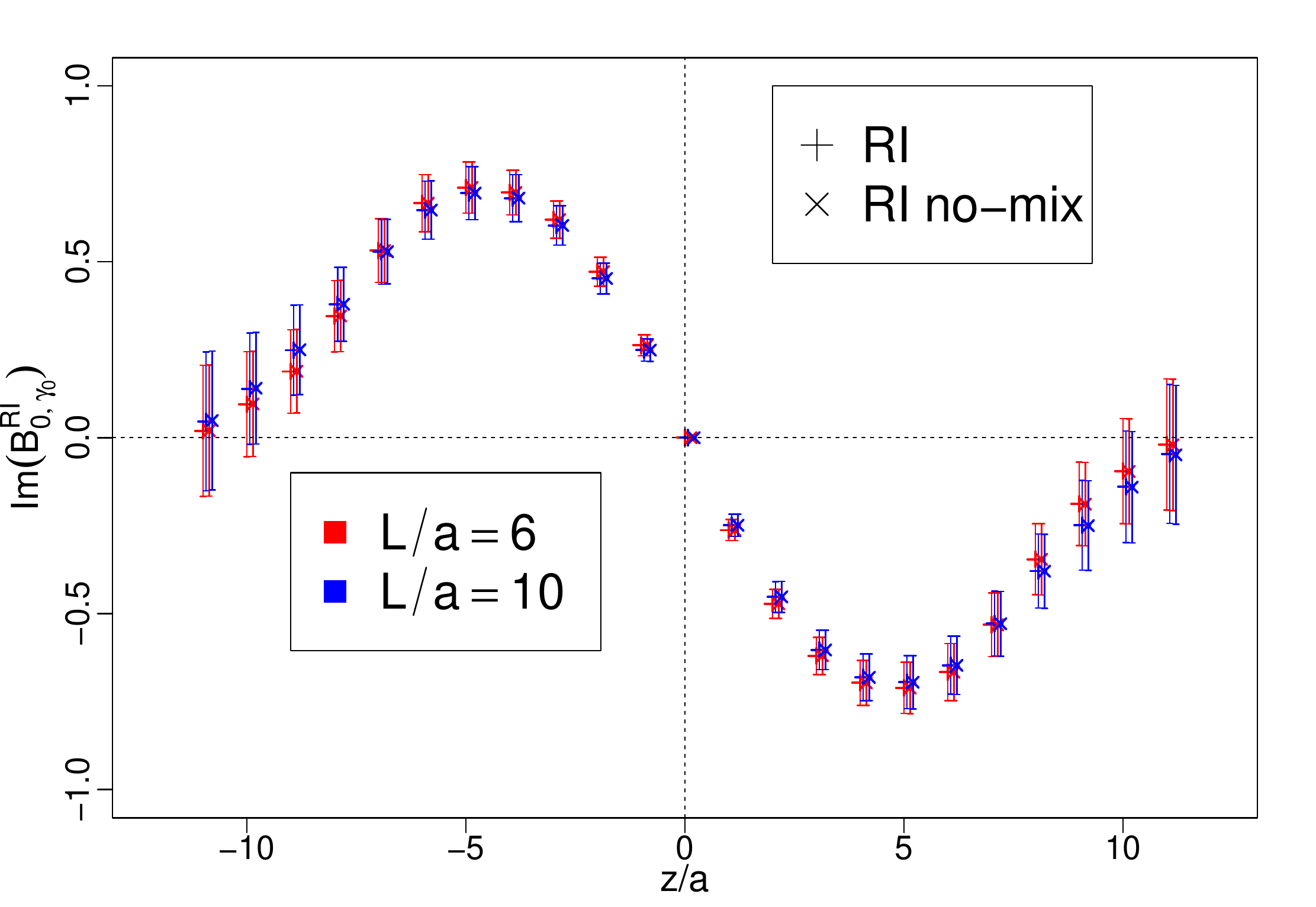}}
	\caption{Real and imaginary parts of the RI/MOM renormalized matrix elements for a transverse separation of $b/a = 1$.}
	\label{fig:B_RI_nomix}
\end{figure}

It is interesting to examine how the relative size of the 
off-diagonal to the diagonal renormalization factors 
changes  as $b$ increases, as the primary focus of lattice 
computation of TMDPDFs is the 
the non-perturbative region in the transverse separation 
$b$. We show in Fig.~\ref{fig:Z_RI_g0_large_b} the 
renormalization factors as a function of $b$ for the 
diagonal and off-diagonal contributions.
We see a contribution of $\lesssim 7-8\%$ up to 
$b = 6a$ for the off-diagonal terms. Although their 
size shows a possible tendency of growth with 
increasing $b$, this growth seems to be only
moderate. We notice that the authors of Ref.~\cite{Shanahan:2019zcq} 
considered the entire set of Dirac structures
for the operator mixing, and observe similar results to
ours, if we take from Ref.~\cite{Shanahan:2019zcq} 
only the results from
the operators that are allowed to mix with $\gamma_0$, 
according to our symmetry arguments.
For example, their result for the 
contribution from $\gamma_5\gamma_1$
is much larger than that of $\gamma_0\gamma_2$ and 
$\gamma_0\gamma_3$, and there is a steady increase going to 
larger $b$ values. For the values of $b$ we use in this 
work, the magnitude of mixing for these operators 
found in Ref.~\cite{Shanahan:2019zcq} 
is also comparable to our findings.
However, considering the dominant diagonal renormalization factors, 
it seems that there are remaining
divergences related to large valus of $z$ and  $b$, as 
also observed in Ref.~\cite{Zhang:2022xuw}. Combined with 
the fact that the bare matrix element decays 
exponentially with $b$~\cite{Ji:2020ect}, it is significantly hard 
to renormalize for large transverse separations using
this scheme. However, in the small $b$ region where 
we can extract information using RI/MOM, it can be 
confirmed that the mixing is negligible. As an example, 
we show in Fig.~\ref{fig:B_RI_nomix} the renormalized 
beam function for the $b=1a$ case, with $+$'s denoting 
the RI/MOM procedure using the full mixing matrix and with the 
$\times$'s denoting the case of only taking into account 
the diagonal contribution. Considering that including 
the mixing has a negligible effect, one can safely ignore 
operator mixing for these values of $b$.
%
%
%
%
%

\subsection{The short distance ratio scheme}
\label{section:Ratio_scheme}

According to Figs.~\ref{fig:Z_RI_g0_off} 
and~\ref{fig:Z_RI_g0_large_b}, mixing with 
different operators can be neglected at least up to
$b/a \lesssim 6$, and we can assume 
that the renormalization of the staple-shaped link is  
multiplicative.
This justifies the approach taken in 
Refs.~\cite{Zhang:2022xuw,LPC:2022zci}.

\begin{figure}[h]
    \centering
    \subfigure{\includegraphics[width=0.49\textwidth]{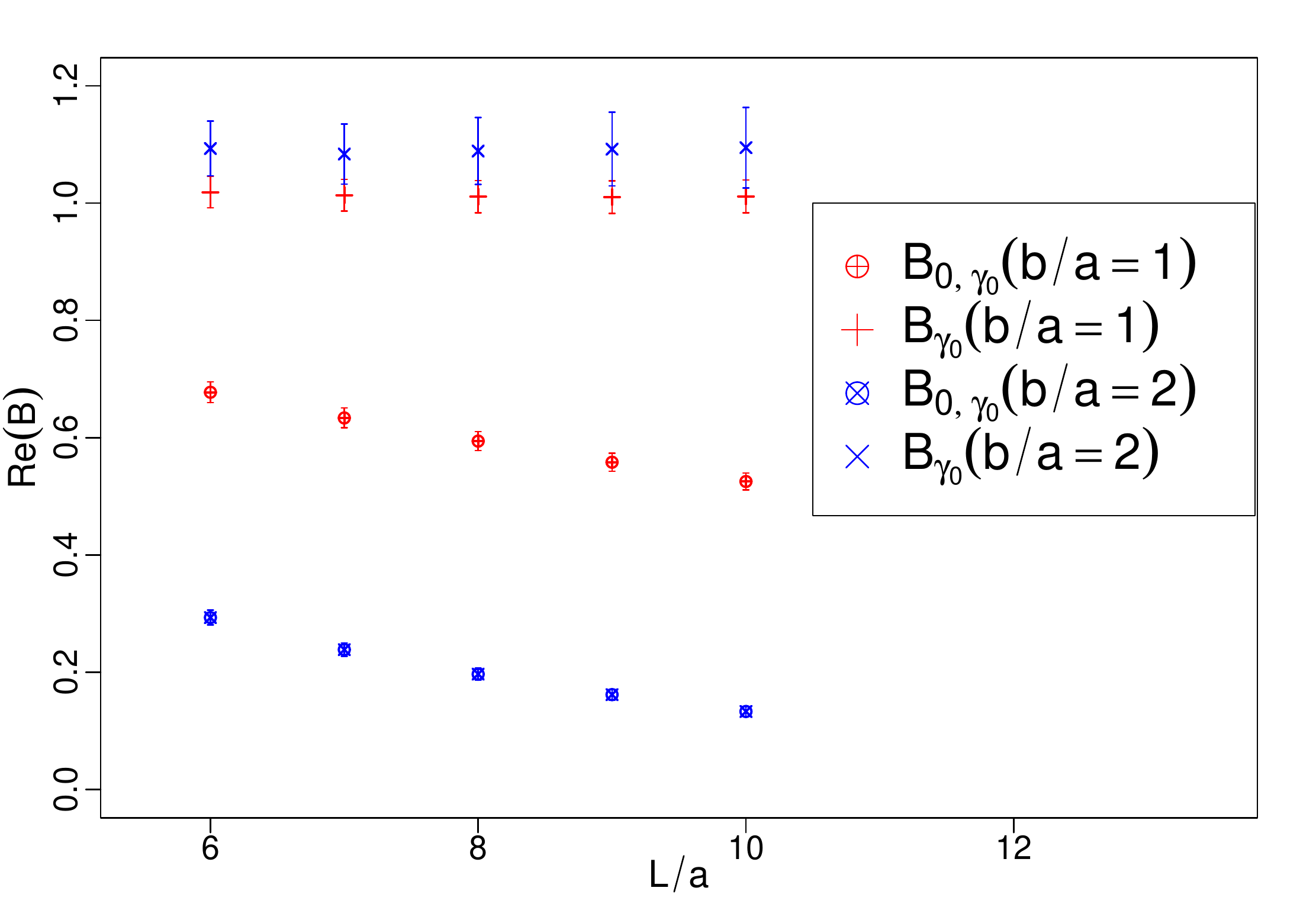}}
    \subfigure{\includegraphics[width=0.49\textwidth]{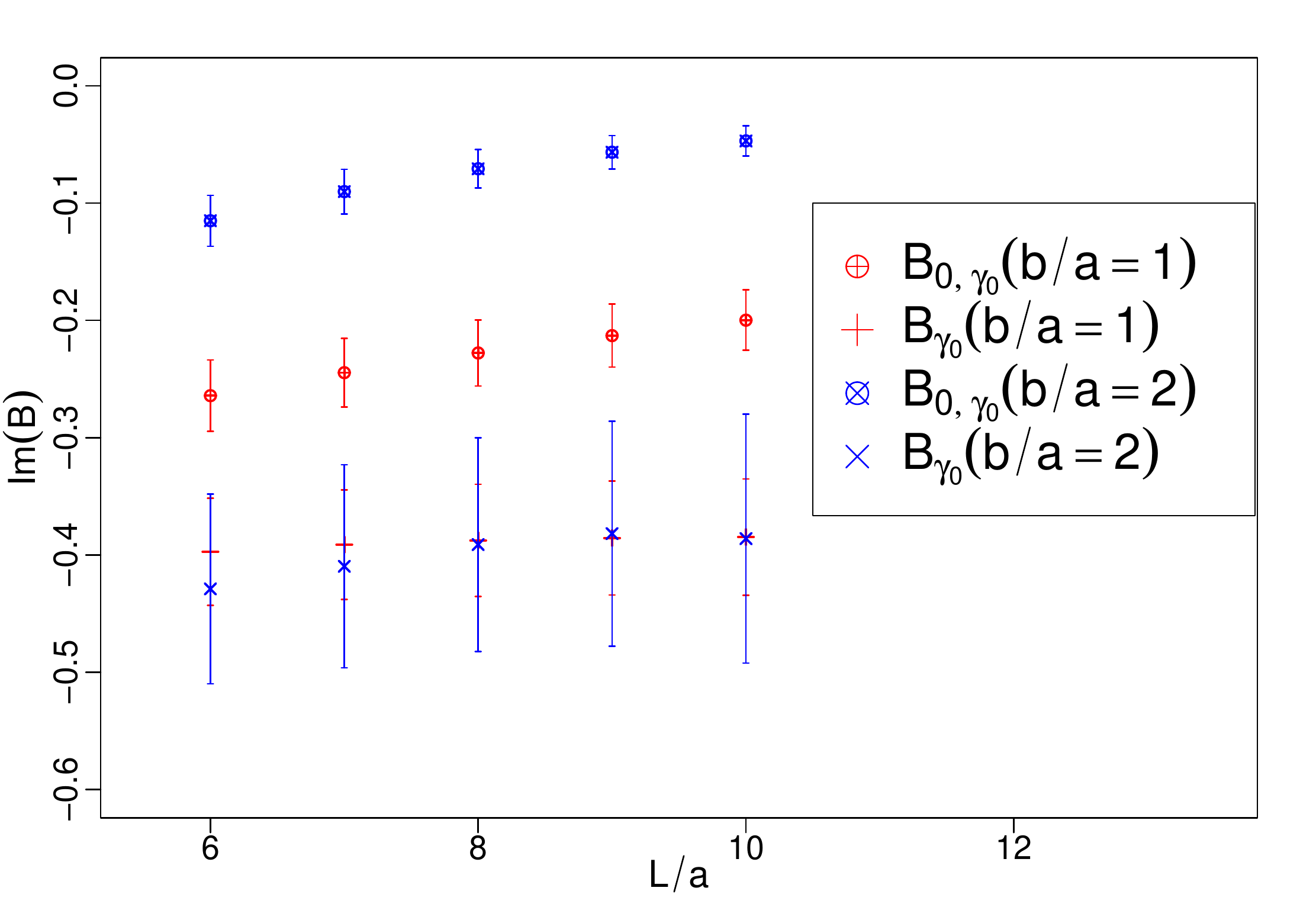}}
    \caption{The effect of taking the ratio  of the bare matrix elements with $\sqrt{Z_E}$ at a fixed $z/a = 2$.}
    \label{fig:B_ZE_sub}
\end{figure}

We first note that the vacuum expectation value of a rectangular
Wilson loop $Z_E$ with sides $2L+z$ and $b$,
\begin{equation}
    Z_E(b,2L+z;1/a) = \frac{1}{3}\Tr\langle 0|{\cal W}(b;2L+z)W_\perp(x+b;b)|0\rangle ,
\end{equation}
is, by construction, a product of the staple-shaped gauge link, as defined 
in Eq.~(\ref{eq:staple_shaped_GL}), and its reflection.
Therefore, $\sqrt{Z_E}$ has the same divergences as that of the staple-shaped gauge link.
Thus, it
should cancel the pinch-pole singularity associated with the length
$L$ of the staple, as well as the divergences associated with the cusps.
Furthermore, as the sides of $Z_E$ are also dependent on the longitudinal
displacement $z$, the exponential divergence associated with 
$z$ present in the staple-shaped link must also be cancelled if 
an appropriated ratio is taken, namely the one proposed in Refs.~\cite{Ji:2019sxk,Zhang:2022xuw}:
\begin{equation}
      B_\Gamma(z,b,P^z;1/a) = 
    \lim_{L\rightarrow\infty}\frac{B_{0,\Gamma}(z,b,L,P^z;1/a)}{\sqrt{Z_E(b,2L+z;1/a)}} 
\end{equation}
To illustrate this point, we show in Fig.~\ref{fig:B_ZE_sub} the 
beam function as a function of $L$, for a fixed $z=2a$ and two values of $b$, 
before and after dividing by $\sqrt{Z_E}$. 
The cancellation of the divergences related to the length of the 
Wilson line is explicitly shown. This ratio, thus, takes care of the divergence associated
with the length $L$ and width $b$ of the staple-shaped operator.\\
After dividing the beam function by the square root of $Z_E$, 
the only remaining divergences are the UV divergences 
associated with the quark field and its end-points 
connecting to the gauge links. As such, they can be cancelled 
by taking ratios, as they have a multiplicative nature~\cite{Ebert:2019tvc,Green:2020xco}.
Hence, one can define
\begin{equation}
    B^{SDR}_\Gamma (z,z_0,b,b_0,P^z)=Z^{SDR}(z_0,b_0;1/a)\,B_\Gamma(z,b,P^z;1/a)\,,
\end{equation}
where,
\begin{equation}
    Z^{SDR}(z_0,b_0;1/a) = \frac{1}{B_\Gamma(z=z_0,b=b_0,P^z=0;1/a)}\,.
\end{equation}
Because the remaining divergences are independent of the length
of the Wilson line, one is free to choose $z_0$ and $b_0$. 
In order to connect these quantities to the $\MSbar$ scheme
via a perturbative scheme conversion, $z_0$ and $b_0$ should be 
small enough for perturbation theory to be valid. 
However, the use of small values for both $z_0$ and $b_0$ 
can introduce sizable discretization errors in the renormalization 
factors, which can affect the validity of the SDR scheme. 
To address this issue, different approaches can be employed in 
order to reduce finite lattice-spacing errors from the non-perturbative 
data for all values of the staple lengths $b/a$ and $z/a$. 
Ideally, the elimination of discretization errors requires 
calculations of physical matrix elements at different finite 
values of the lattice spacing $a$ and an extrapolation $a\rightarrow 0$. 
When data for multiple values of $a$ are not available, a number 
of different approaches can be employed in order to reduce 
discretization errors at each lattice spacing. A standard 
method is to apply an improved discretization, in both 
the action and the operators under study, using the 
Symanzik-improvement program~\cite{Symanzik:1983gh,Symanzik:1983dc}. 
Another way to reduce this kind of systematic error from 
a lattice calculation is to subtract one-loop artifacts 
employing lattice perturbation theory to all orders in $a$,
from the non-perturbative vertex functions calculated in lattice 
simulations. Our group has successfully applied this method 
to the renormalization of local quark bilinear operators~\cite{Constantinou:2009tr,Constantinou:2013ada,Alexandrou:2015sea}, 
and more recently to the renormalization of non-local straight 
Wilson-line operators for quasi-PDFs~\cite{Constantinou:2022aij}. 
These studies have provided a useful feedback on the effectiveness 
of artifacts in the renormalization factors for different ranges of 
the scales entering the renormalization procedure. Since this is 
our first non-perturbative study considering non-local staple-shaped 
operators we do not consider finite-$a$ errors, but we intend to 
apply the method of
subtracting one-loop artifacts in a future extension of our study.

The renormalized matrix elements are converted to the $\MSbar$
scheme using perturbation theory. We have computed the vertex, sail, and
tadpole one-loop diagrams for external quark states with a general momentum
$p^z$. For $p^z\rightarrow 0$, we obtain
\begin{equation}
    Z^{\MSbar,SDR}(z_0,b_0,\mu_0)=1+\frac{\alpha_sC_F}{2\pi}\left(\frac{1}{2}+\frac{3}{2}\ln\left(\frac{b_0^2+z_0^2}{4e^{-2\gamma_E}}\mu_0^2\right)-2\frac{z_0}{b_0}\textrm{arctan}\frac{z_0}{b_0}\right)\,,
\end{equation}
which agrees with Eq.~(6) of Ref.~\cite{Zhang:2022xuw}. 
Note that this factor equals  
$B_\Gamma^{\MSbar} (z=z_0, b=b_0, P^z = 0,\mu_0)$. Details of this
calculation for a general external momentum will be presented 
in Ref.~\cite{Constantinou:2023}. The renormalized beam function in the $\MSbar$ scheme
is then given by
\begin{equation}
    B^{\MSbar}_\Gamma (z,b,P^z)=Z^{\MSbar}(z_0,b_0,\mu_0) B^{SDR}_\Gamma(z,z_0,b,b_0,P^z).
\label{eq:msbar_Ratio_conversion}
\end{equation}
Results for $B^{\MSbar}(z,b,P^z)$ are presented in 
Section~\ref{section:results}, where we  also discuss 
the  cancellation of the $z_0,b_0$ dependence in Eq.~(\ref{eq:msbar_Ratio_conversion}).

\subsection{Short distance RI/MOM}
\label{section:short_RI/MOM}

As discussed in Section~\ref{section:RI/MOM}, the usual 
$\RI$ scheme may be problematic at large $z$ and $b$ as the 
magnitude of the $Z^{\RI}$ factors grows exponentially. 
Also, as shown in Refs.~\cite{Zhang:2020rsx,Zhang:2022xuw}, 
the usual $\RI$ scheme may still contain a residual 
linear divergence, which may not be properly canceled. 
On the other hand, as discussed in Section \ref{section:Ratio_scheme}, 
the $\sqrt{Z_E}$ factor cancels all divergences in $z$ and $b$ 
present in the staple. Hence, we can define a vertex function 
that is free of such divergences,
\begin{equation}
    \Lambda^\Gamma(z,b,p;1/a) = \frac{\Lambda^\Gamma_0(z,b,p;1/a)}{\sqrt{Z_E(b,2L+z;1/a)}}.
\end{equation}
Because the divergences related to the lengths of the Wilson
line have been removed, 
we can compute the renormalization factors as in 
Section~\ref{section:Ratio_scheme} at some fixed $z_0,b_0$~\cite{Ji:2021uvr}
\begin{gather}
\frac{Z^{\RI-\text{short}}_{\Gamma\Gamma^{'}}(z_0,b_0,\mu_0;1/a)}{Z^{\RI}_q(\mu_0;1/a)}\frac{1}{12} {\rm Tr} \left[\frac{\Lambda^\Gamma(z,b,p;1/a)\Gamma^{'}}{e^{ip^zz +ib p_\perp}}\right] \Bigg|_{p{=}\mu_0,z=z_0,b=b_0} {=} 1\, . 
\end{gather} 
The $Z^{\RI-\text{short}}$ factor defined at 
fixed $z_0$ and $b_0$ is used to renormalize the bare $B_{\gamma_0}$ 
defined at arbitrary values of $z$ and $b$. We have labeled this
procedure as RI-short.
In principle, the vertex function in the
standard RI/MOM could also be modified by taking its ratio
with $\sqrt{Z_E}$. This would reduce the growth of the
$Z$-factors with increasing $b$. However, when combining $Z^{\RI}$
with the bare $B_{\gamma_0}$, the $\sqrt{Z_E}$ factors cancel
each other, 
and therefore $\sqrt{Z_E}$ has no effect on the
renormalized matrix elements.
On the other hand, the $\sqrt{Z_E}$ factor appearing 
in the vertex function of the RI-short scheme is defined at 
fixed values of $z$ and $b$, contrary to the $\sqrt{Z_E}$ 
factor appearing in the bare $B_{\gamma_0}$. Hence the 
cancellation of the $Z_E$ factors do not happen
in the RI-short scheme.

As in Section~\ref{section:Ratio_scheme}, we choose the 
pair $z_0,b_0$ to be in the perturbative region in order 
to make the perturbative conversion to the $\MSbar$ scheme reliable. 
Moreover, the
study of possible lattice artifacts associated with the use 
of small values of $z_0/a$ and $b_0/a$ will be considered in future 
extensions of our study by using one-loop lattice perturbation theory.

\begin{figure}[!h]
    \centering
    \includegraphics[width=0.6\textwidth]{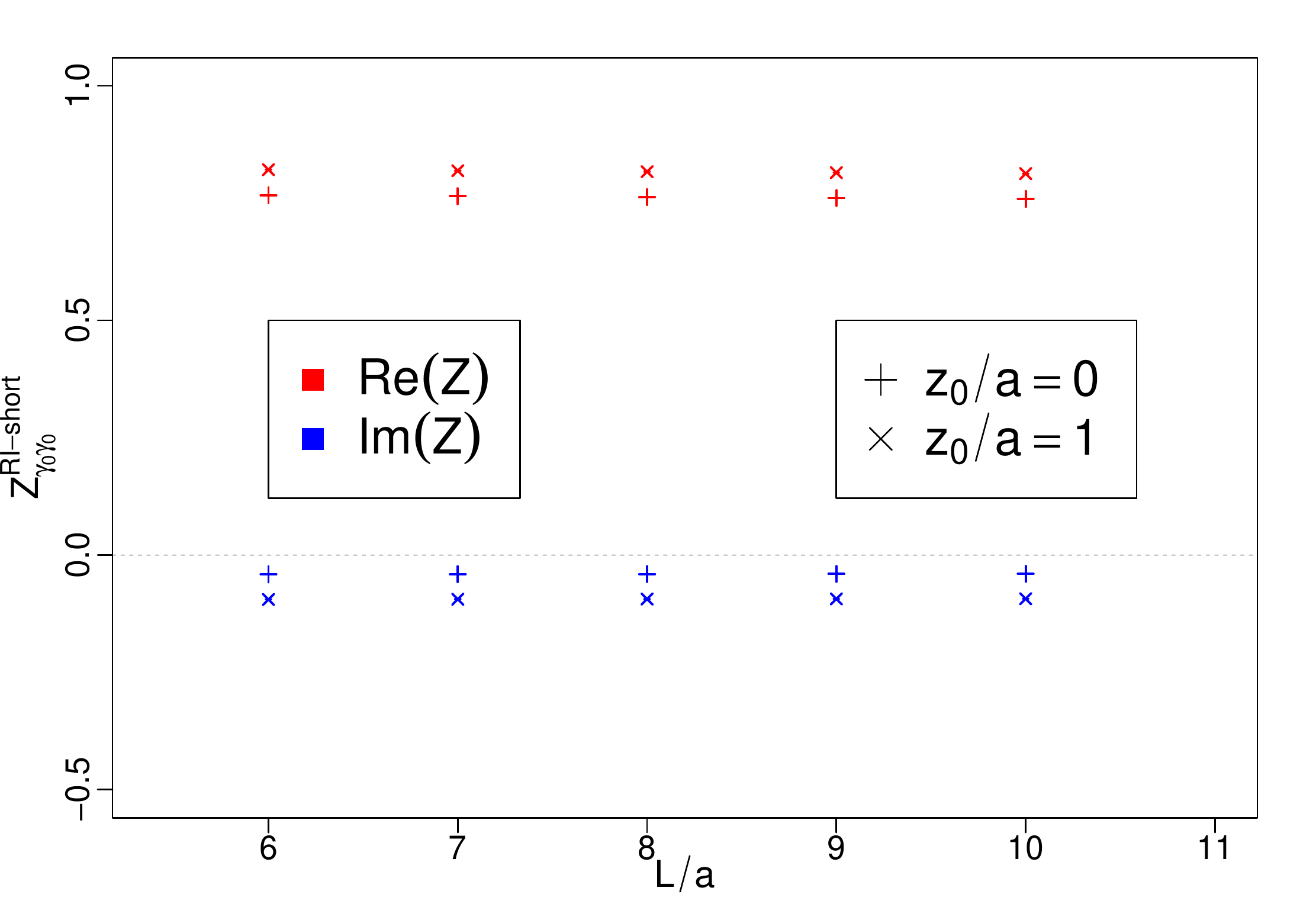}
    \caption{The $Z^{\RI-\text{short}}$ renormalization factor within the perturbative range  
    $b_0/a = 1$ as a function of the length $L$ of the staple.}
    \label{fig:Z_RI_short}
\end{figure}
\begin{figure}[!h]
    \centering
    \includegraphics[width=\textwidth]{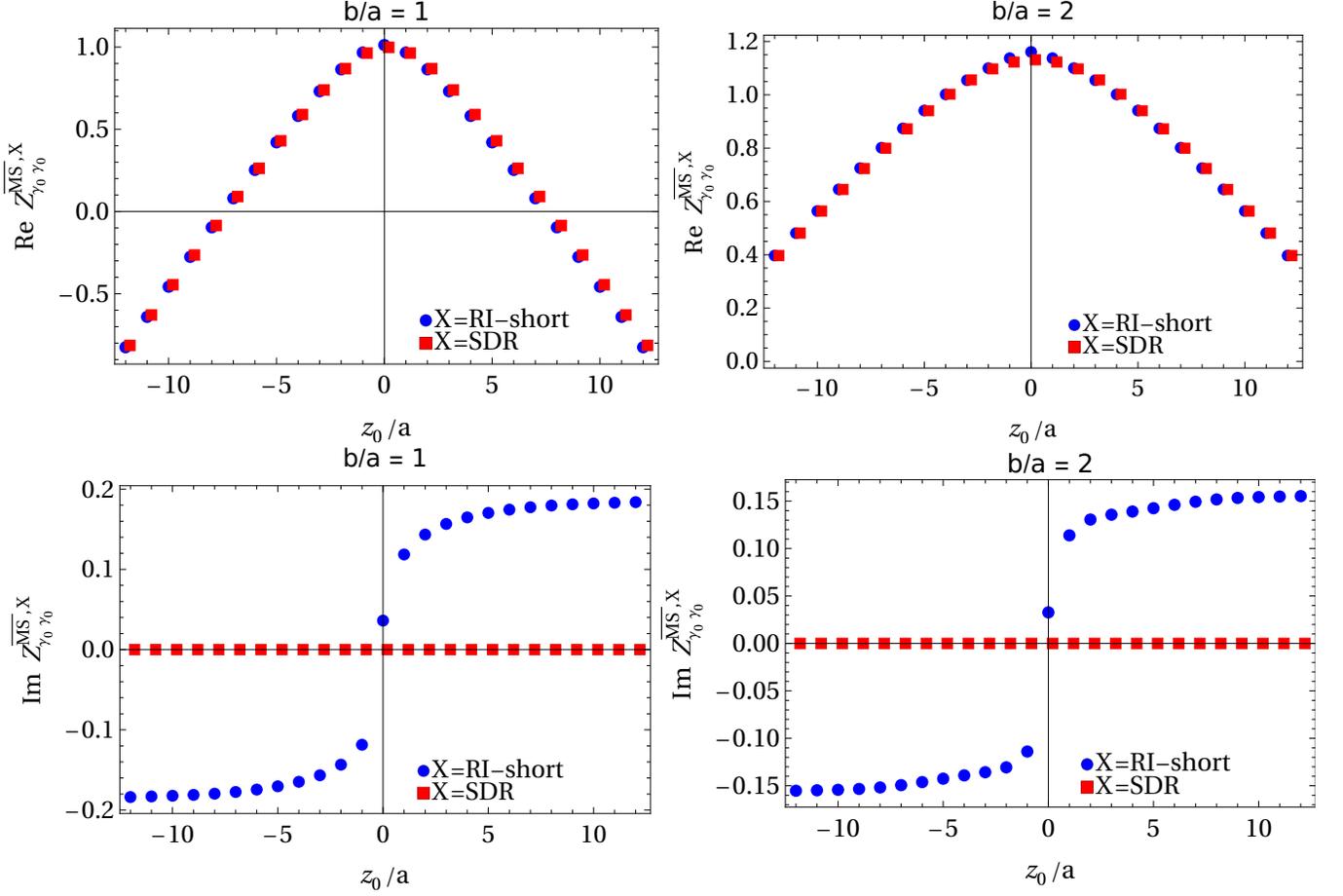}
    \caption{Conversion factor to the $\MSbar$ scheme (top: real part, bottom: imaginary part) for the RI-short and SDR schemes at $b/a = 1$ (left) and $b/a = 2$ (right).}
    \label{fig:conversion}
\end{figure}

The corresponding renormalized beam function in this scheme is then given by
\begin{equation}
    B^{\RI-\text{short}}_\Gamma (z,z_0,b,b_0,\mu_0,P^z)=
    \sum_{\Gamma^{'}}\left[Z^{\RI-\text{short}}_{\cal O}(z_0,b_0,\mu_0;1/a) \right]_{\Gamma\Gamma^{'}}B_{\Gamma^{'}}(z,b,P^z;1/a).
\end{equation}
For illustration, we show in Fig~\ref{fig:Z_RI_short} 
the renormalization factors $Z^{\RI-\text{short}}_{\cal O}(z_0,b_0,\mu_0;1/a)$ 
for $z_0=0,\,1a$, $b_0=1a$ 
and $\mu_0 = \frac{2 \pi}{a} \left( \frac{6 + 0.5}{48}, \frac{3}{24}, \frac{3}{24}, \frac{3}{24}\right)$. 
We also show the off-diagonal 
renormalization factors, which, in principle, mix with $\gamma^0$. 
As expected, they are also independent of $L$ and can be omitted
in the full calculation as their contribution is negligible  
 as compared to the diagonal contribution.
Finally, we convert $ B^{\RI-\text{short}}_\Gamma$ to the $\MSbar$ scheme 
using one-loop perturbation theory,
\begin{equation}
    B^{\MSbar}_\Gamma (z,b,\mu_0,P^z)=
\sum_{\Gamma^{'}}\left[Z^{\MSbar,\RI-\text{short}}_{\cal O}(z_0,b_0,\mu_0) \right]_{\Gamma\Gamma^{'}}B^{\RI-\text{short}}_{\Gamma^{'}}(z,z_0,b,b_0,\mu_0,P^z)\,.
\end{equation}
The conversion matrix $Z_O^{\MSbar,\RI-\text{short}}$ is
calculated in dimensional regularization for arbitrary values 
of the momentum 
scale $\mu_0$. Explicit expressions for all $\Gamma, \Gamma'$ 
are given in Ref.~\cite{Constantinou:2023}. 
Similar perturbative studies can be found in Refs.~\cite{Constantinou:2019vyb,Ebert:2019tvc}. A comparison
between the conversion factor in RI-short and the SDR scheme 
is shown in Fig.~\ref{fig:conversion} for the case 
$\Gamma = \Gamma' = \gamma_0$. 
We observe that the real parts are compatible between the two 
schemes, while the RI-short scheme shows a non-zero imaginary part, 
in contrast to the SDR scheme. This is a consequence of using a
non-zero momentum scale 
$\mu_0$ in the RI-short scheme, while SDR is defined at zero momentum. 
Note that specific choices of $\mu_0$ in RI-short can lead to a vanishing imaginary 
part in the renormalization factors. In particular, by setting to zero the two 
momentum components parallel to the staple segments, the one-loop expression 
for the renormalization factors results into a vanishing imaginary part. 
However, such choice of momentum gives rise to unwanted Lorentz non-invariant 
contributions in the non-perturbative calculations. Thus, in our study we 
follow the common practice of employing democratic momentum with reduced Lorentz 
non-invariant contributions at the cost of introducing imaginary part in the 
renormalization factors.

\section{Results}
\label{section:results}

In this Section, we present the renormalized beam functions in
the 3 renormalization schemes discussed in 
Section~\ref{section:non_pert_renom}, as well as the 
corresponding results in the $\MSbar$ scheme.

\begin{figure}[h]
	\centering
	\subfigure{\includegraphics[width=0.49\textwidth]{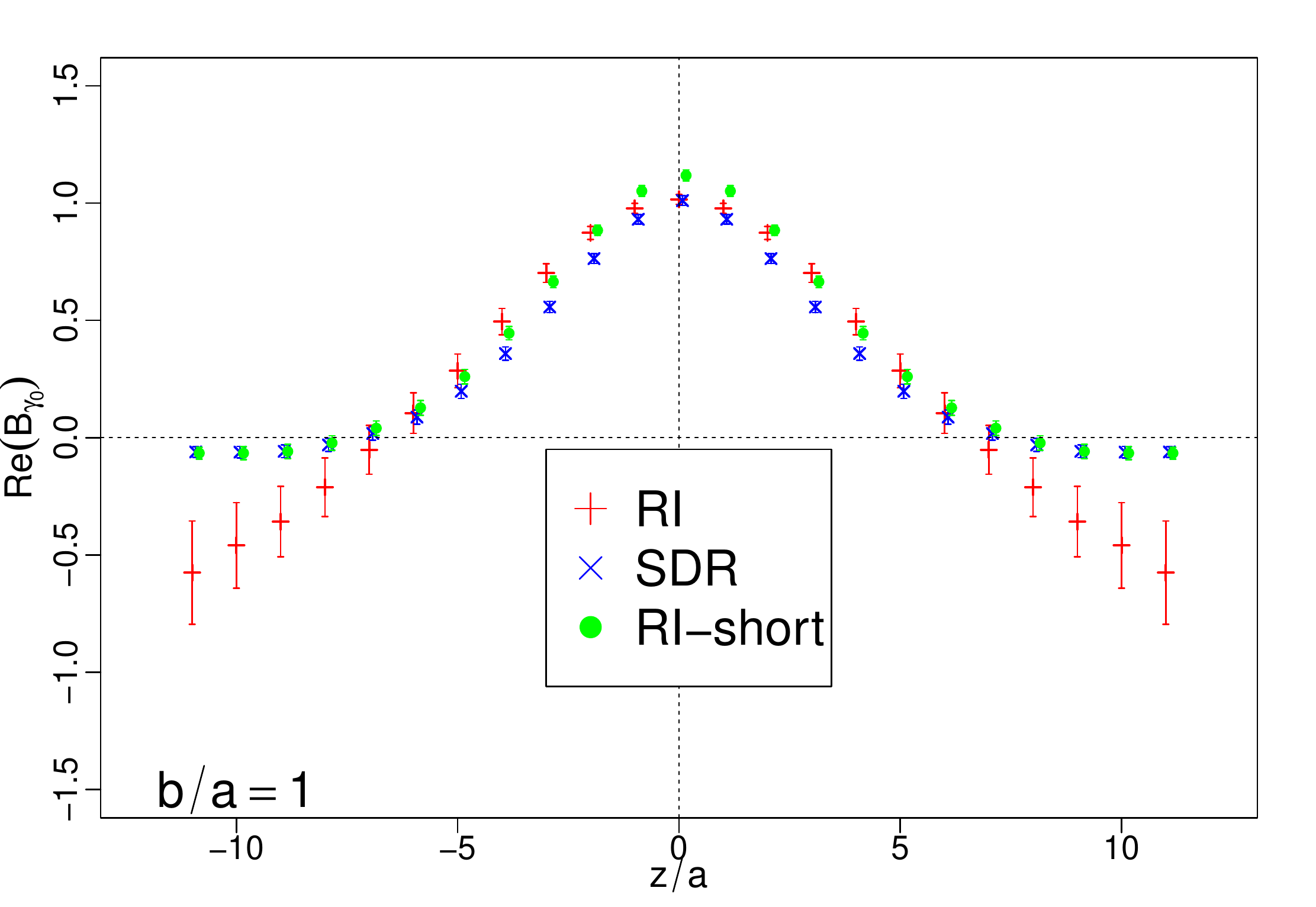}}
	\subfigure{\includegraphics[width=0.49\textwidth]{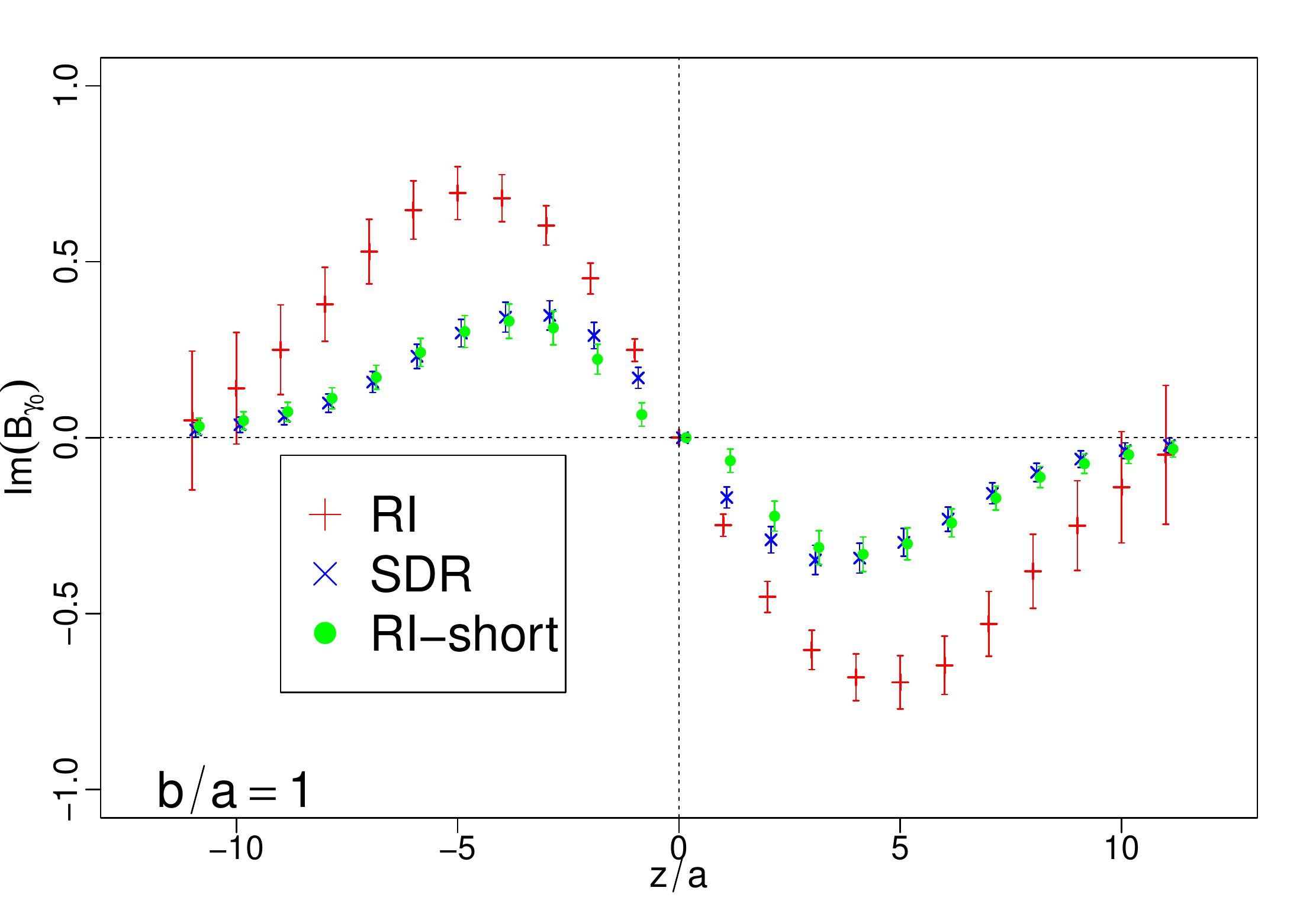}}\\
        \subfigure{\includegraphics[width=0.49\textwidth]{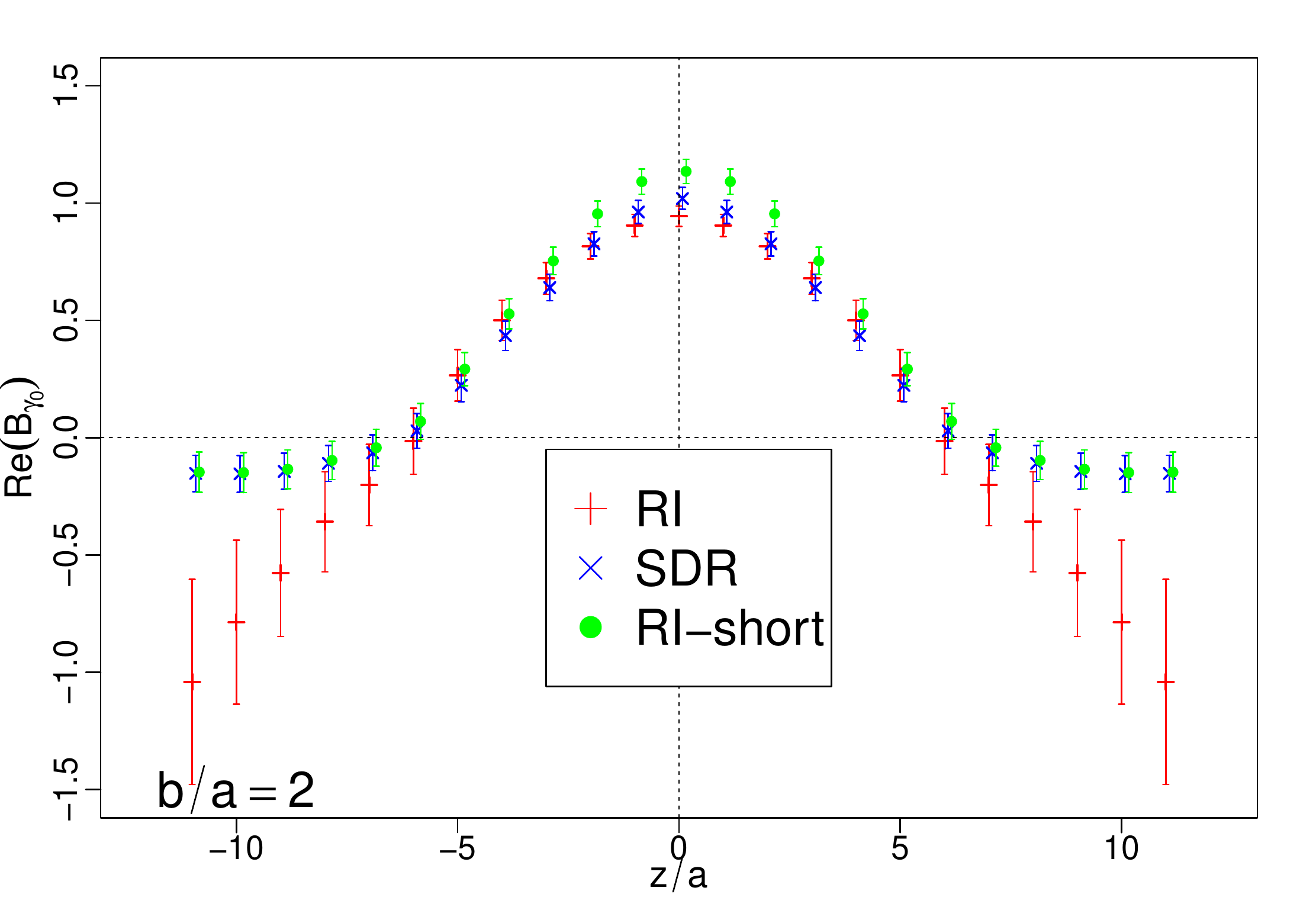}}
	\subfigure{\includegraphics[width=0.49\textwidth]{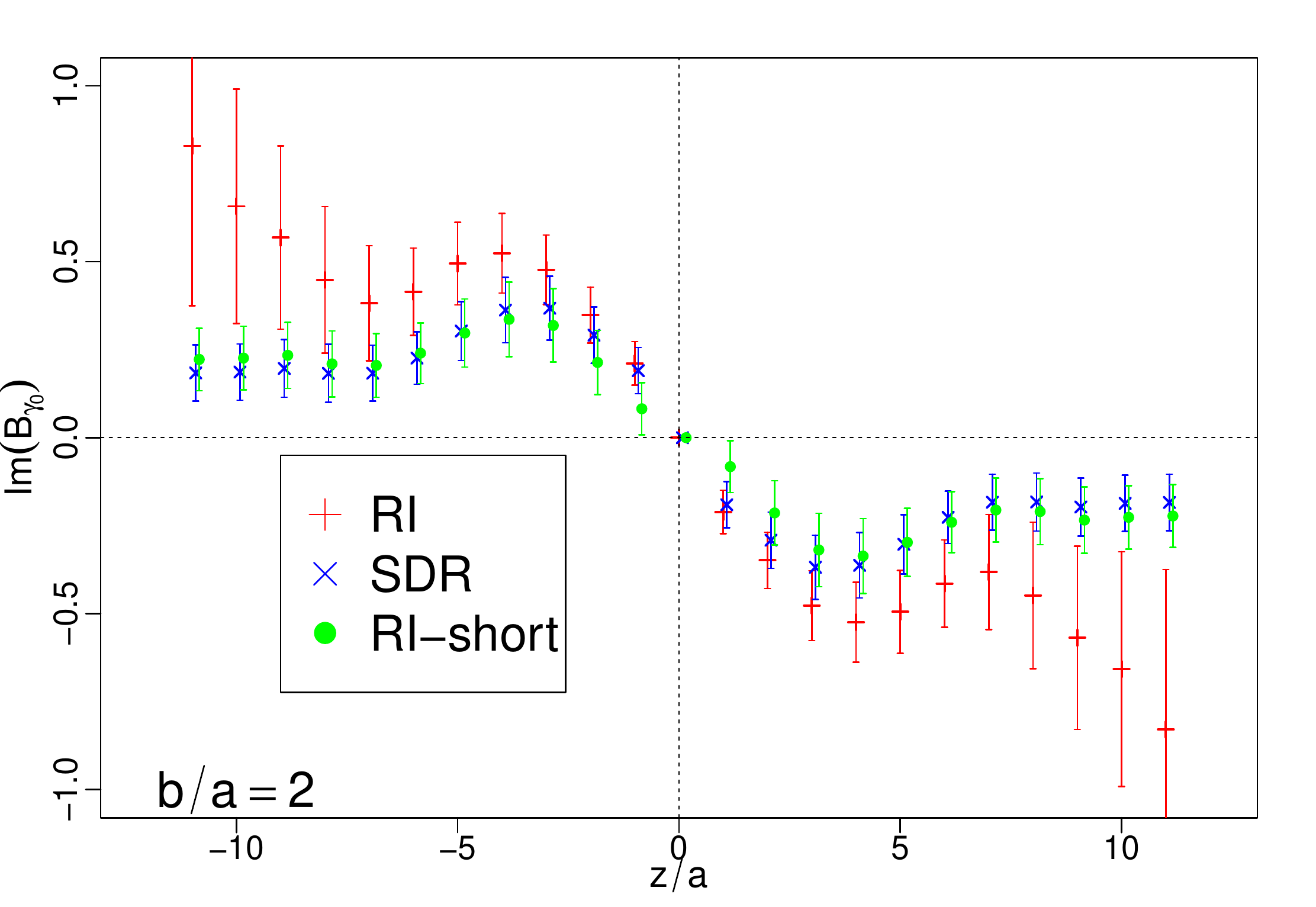}}
	\caption{Renormalized matrix elements for $z_0/a=b_0/a=1$
             for the SDR and the $\RI-\text{short}$ schemes, and $P^z\simeq 1.7$ GeV.}
	\label{fig:B_renorm}
\end{figure}

In Fig.~\ref{fig:B_renorm}, we show the real and imaginary parts
of $B_{\gamma_0}(z,b,\mu_0,P^z)$ as a function of $z$ for two values
of the transverse separation, $b/a=1,2$. For the real parts, 
the schemes agree within errors for $|z/a|\lessapprox 6$. 
For $|z/a|\gtrapprox 6$, $B^{SDR}_{\gamma_0}$ and
$B^{\RI-\text{short}}_{\gamma_0}$ are consistent, 
while $B^{\RI}_{\gamma_0}$ becomes increasingly
more negative with increasingly larger errors. 
This discrepant behaviour between the RI-MOM and 
the RI-short and SDR schemes indicates the existence of 
a residual linear divergence at large $|z|$ in the standard RI/MOM,
as observed in~\cite{Zhang:2022xuw}.
For the imaginary parts, we observe a similar behavior. 
Namely, the SDR and the $\RI$-short schemes are almost 
identical for any $|z/a|$ for the two values of $b/a$ considered.
For the usual RI/MOM, we observe deviations not only for
$|z/a|\gtrapprox 6$ but also for most of the $|z/a|$ region 
with an increase in errors with increasing $|z/a|$.

\begin{figure}[h]
    \centering
    \subfigure{\includegraphics[width=0.49\textwidth]{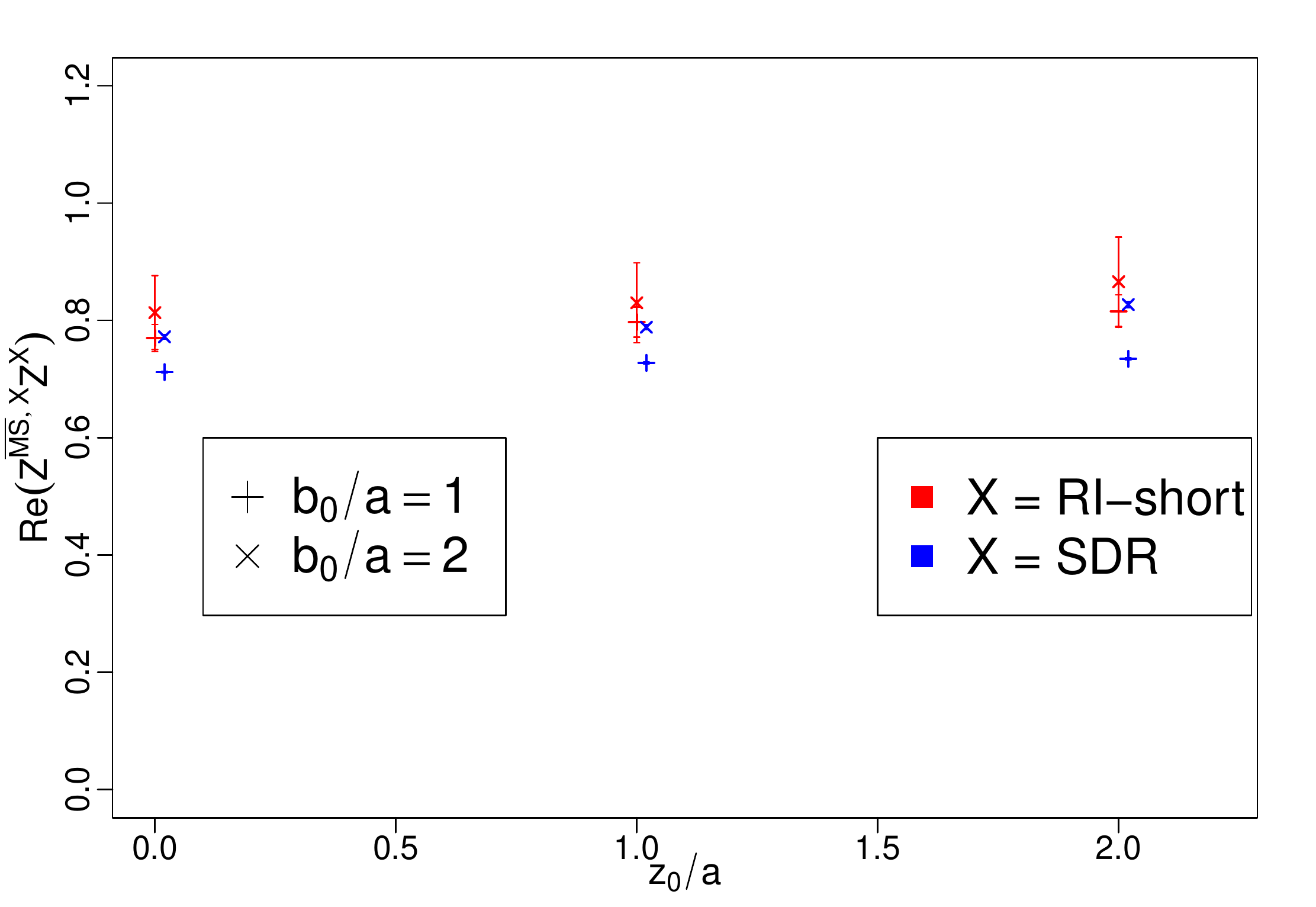}}
    \subfigure{\includegraphics[width=0.49\textwidth]{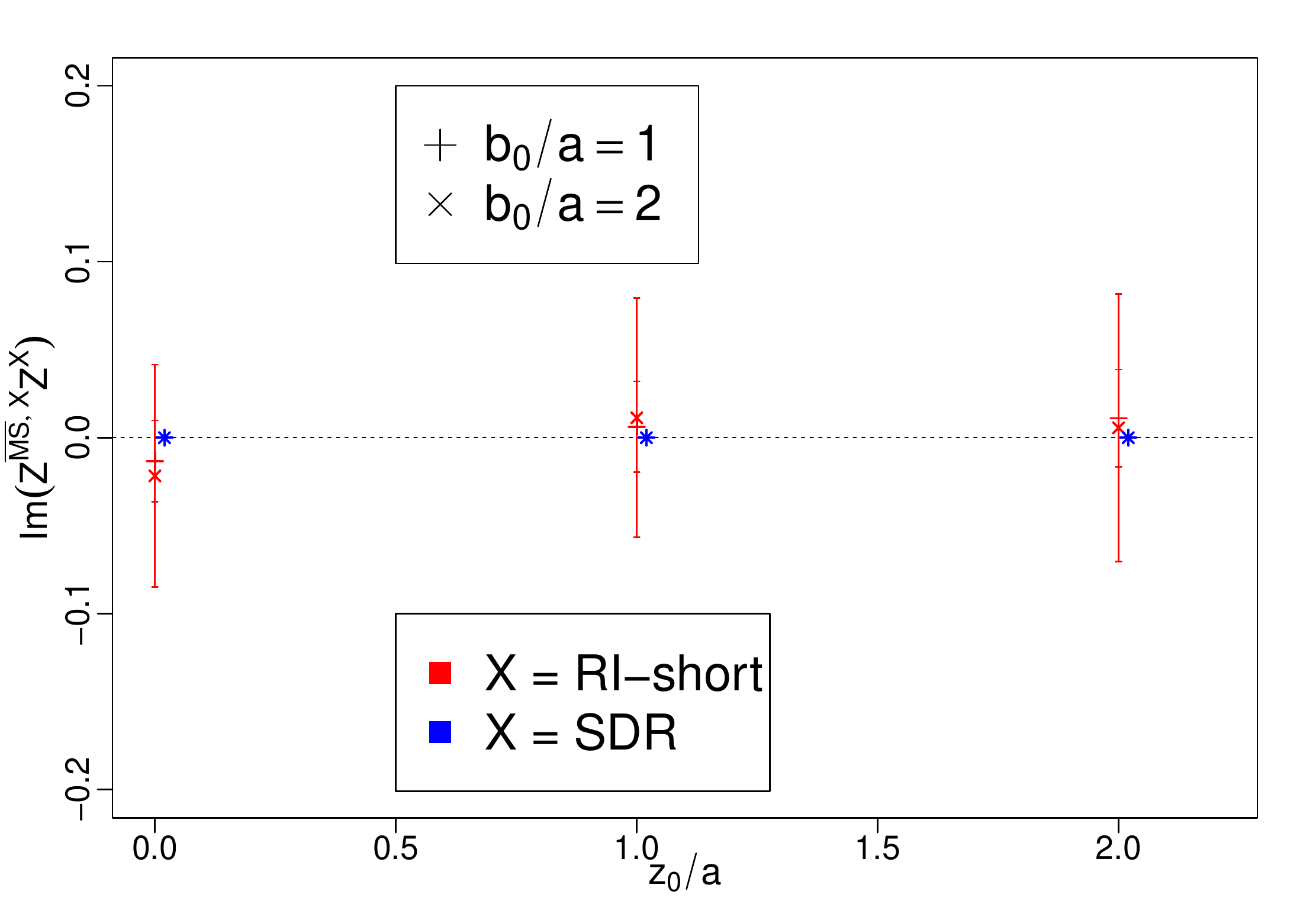}}
    \caption{Dependence of the product between the $\MSbar$ conversion factors and the SDR and $\RI-\text{short}$ renormalization factors on the choice of the short perturbative scale.}
    \label{fig:Z_MS_comparison}
\end{figure}

\begin{figure}[h]
	\centering
	\subfigure{\includegraphics[width=0.49\textwidth]{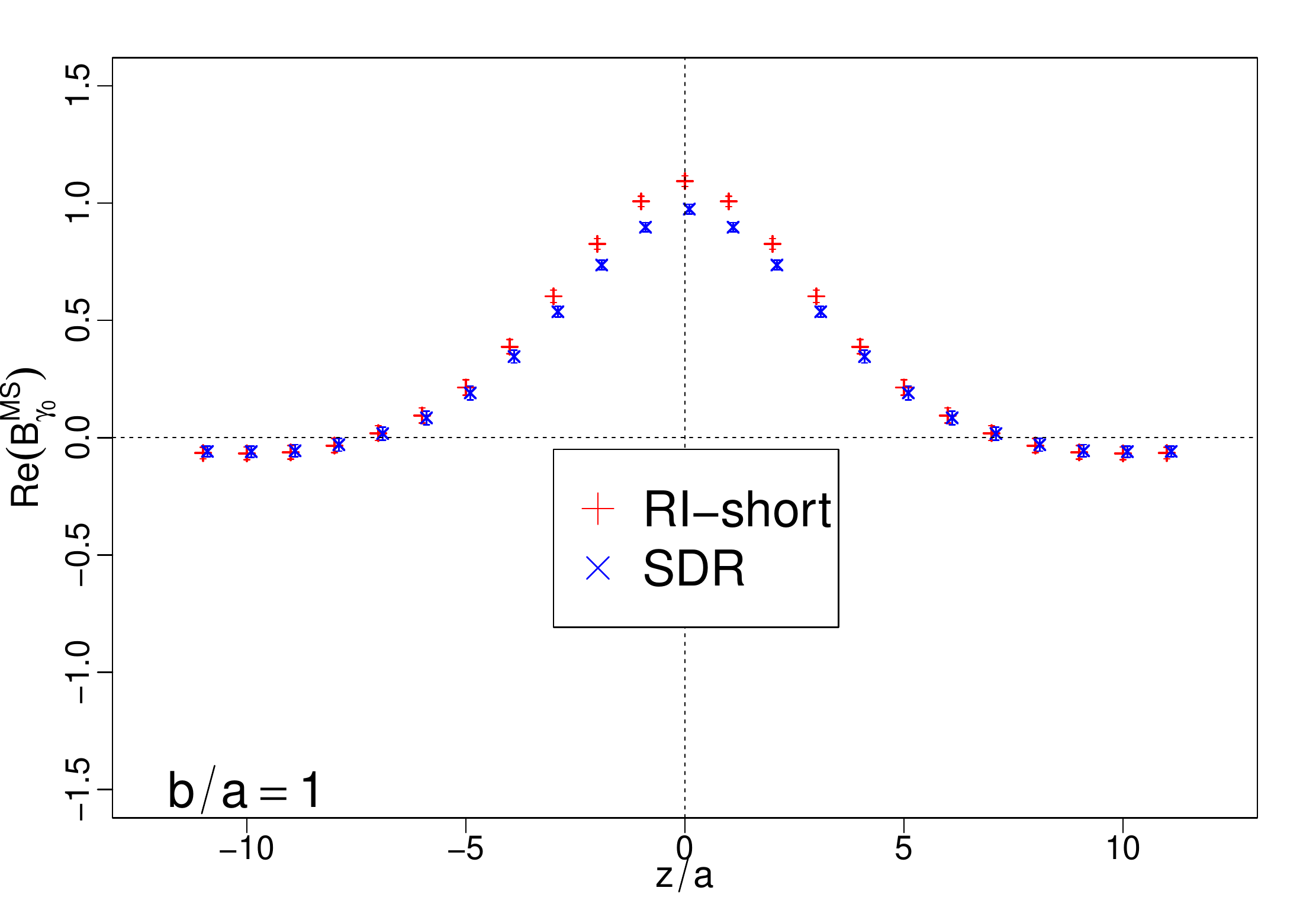}}
	\subfigure{\includegraphics[width=0.49\textwidth]{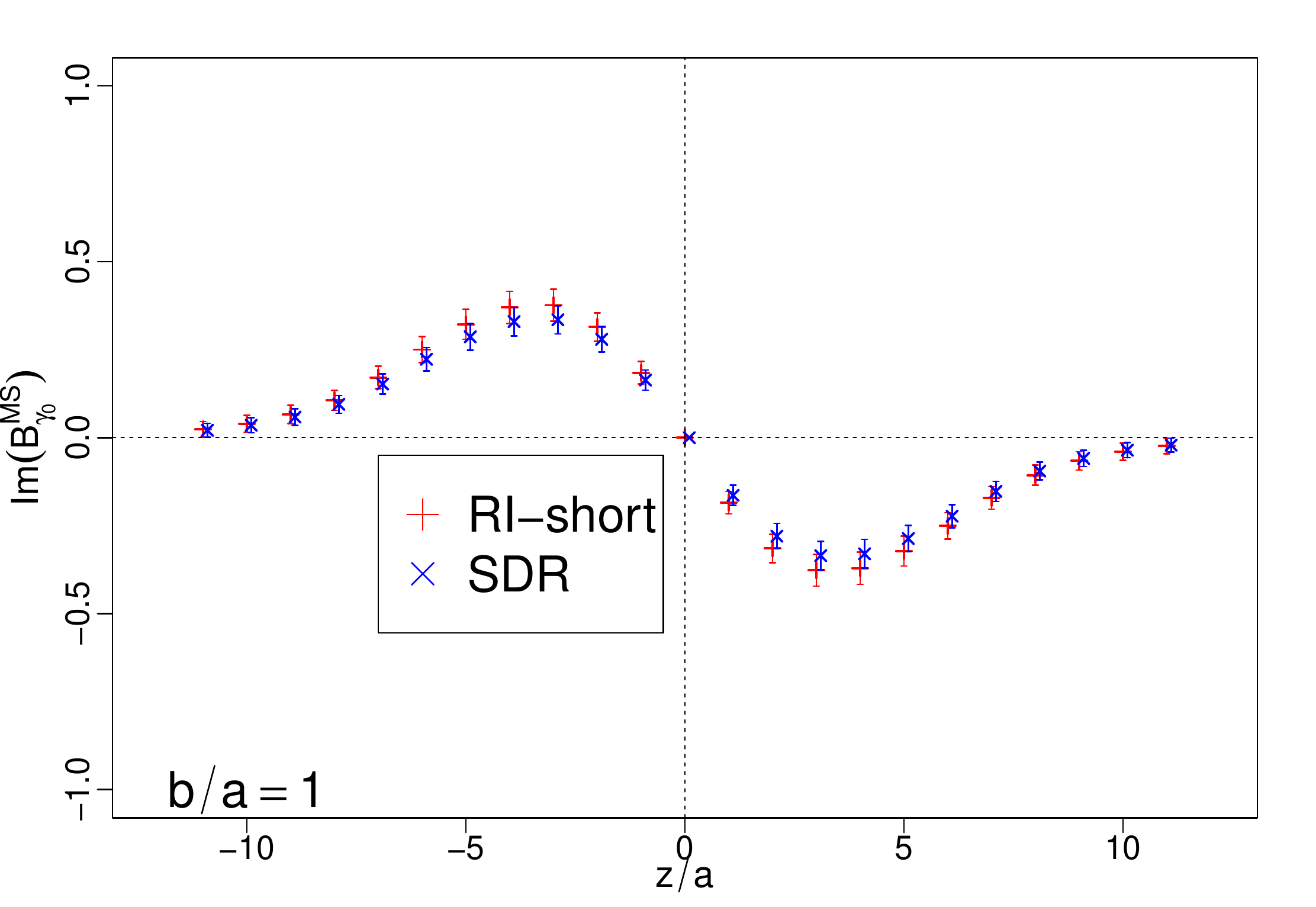}}\\
        \subfigure{\includegraphics[width=0.49\textwidth]{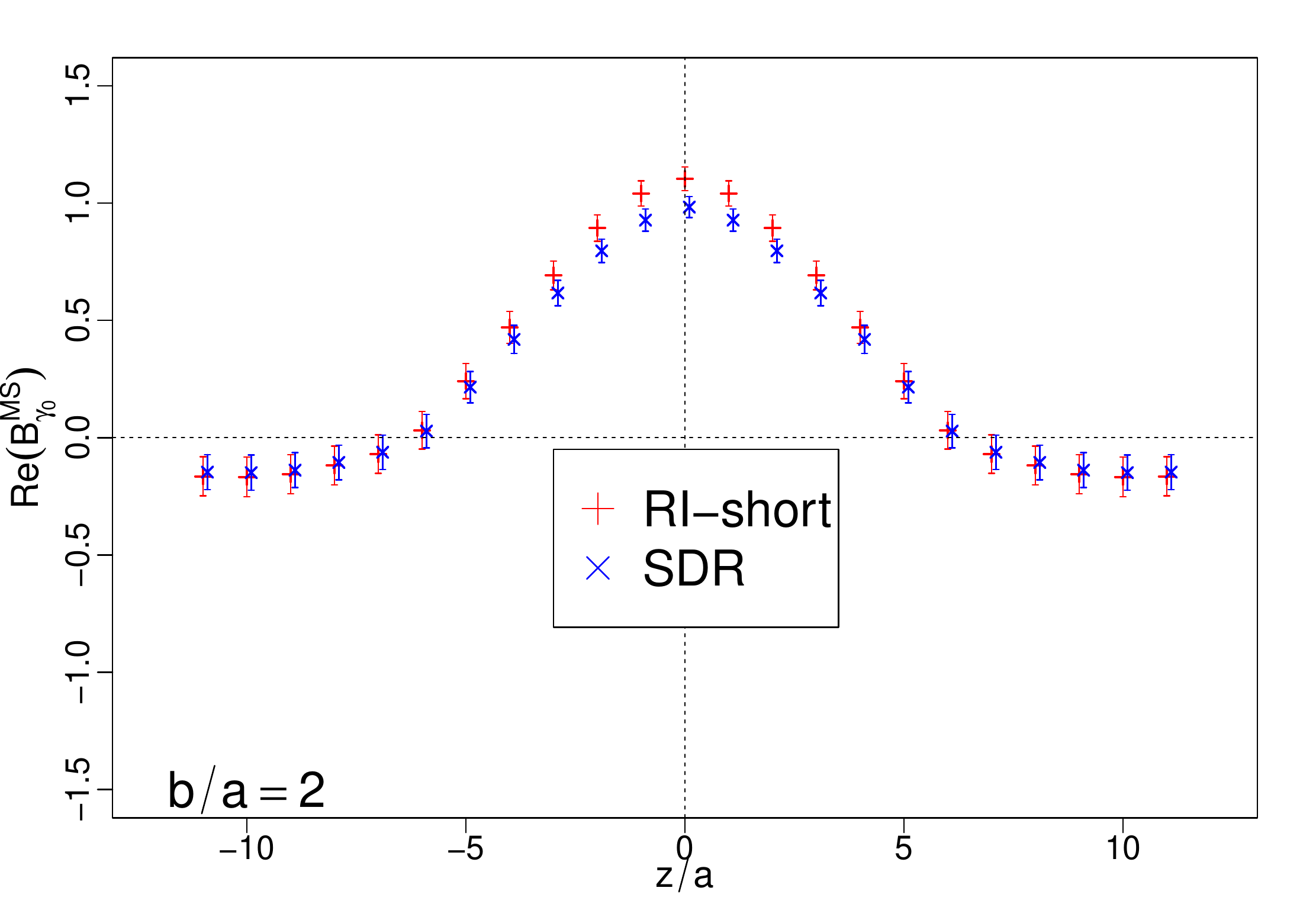}}
	\subfigure{\includegraphics[width=0.49\textwidth]{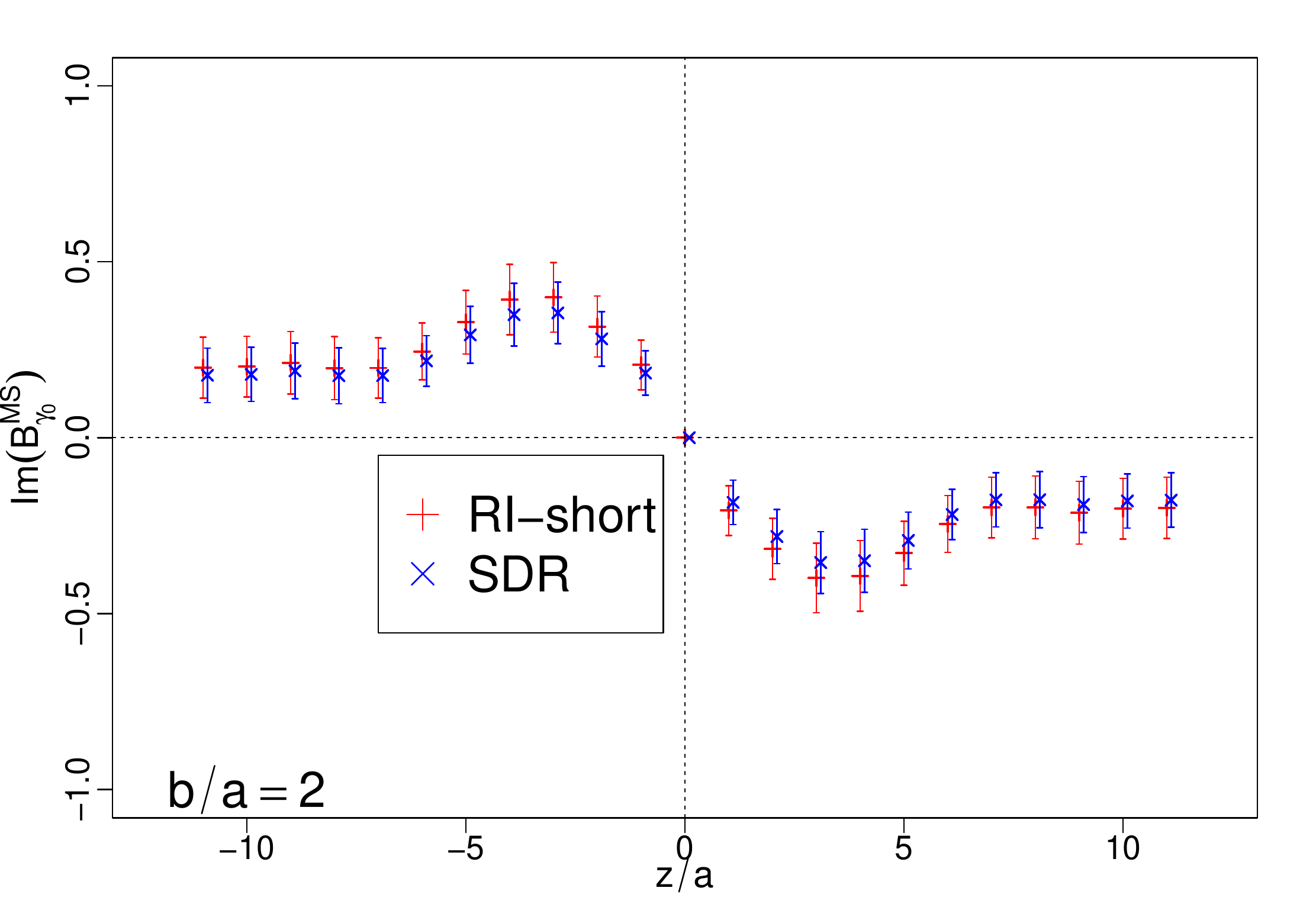}}
	\caption{$\MSbar$ renormalized matrix elements converted from the SDR and the 
             $\RI-\text{short}$ schemes, using $z_0/a=b_0/a=1$ and $P^z\simeq 1.7$ GeV}
	\label{fig:B_renorm_MSbar}
\end{figure}

The conversion  to the $\MSbar$ scheme is 
done using one-loop perturbation theory for 
all three schemes considered. For the two cases where 
the renormalization factors are fixed at a 
short perturbative scale, we use $z_0/a=b_0/a=1$ 
when computing $B^{\RI-\text{short}}_{\Gamma^{'}}(z,z_0,b,b_0,\mu_0,P^z)$ 
and $B^{SDR}_\Gamma(z,z_0,b,b_0,P^z)$. Since 
the final results in the $\MSbar$ scheme should 
be independent of the values we choose for the 
$z_0,b_0$ pair, we examine the stability of our results 
on the choice of the values for this 
pair. To this end, we show in Fig.~\ref{fig:Z_MS_comparison} 
the product of the renormalization factors for 
each scheme with the corresponding conversion 
factors to the $\MSbar$ scheme, $Z^{\MSbar,SDR}_{\gamma_0}Z^{SDR}_{\gamma_0}$ 
and $Z^{\MSbar,\RI}_{\gamma_0}Z^{\RI-\text{short}}_{\gamma_0}$. 
These products should be independent on the 
choice of values of  $z_0,b_0$ as long as the one-loop 
perturbative calculation is a reliable approximation. 
We notice that these products are equal to the renormalization
factors in the $\MSbar$ scheme in the absence of mixing.
For the ratio scheme, there is a $10\% - 20\%$ correction 
when going from $b_0/a=1$ to $b_0/a=2$ for the three 
values of $z_0$ considered here, with the correction
increasing for larger values of $z_0$.
This indicates a limitation on the use of one-loop 
perturbation theory to perform the conversion to 
the $\MSbar$ scheme already at a transverse separation 
as low as $b/a=2$. As a consequence, $\mathcal{O}(\alpha_s^2)$ 
corrections cannot be disregarded. For the $z_0$ dependence, 
the choice of  $z_0/a=0$ and $z_0/a=1$ is nearly 
equivalent within one sigma error. For $z/a=2$,
the discrepancy grows, especially for $b/a=2$. 
For the RI-short scheme, the situation is not 
significantly changed. From this discussion, we 
conclude that our choice for the pair $z_0,b_0$ is 
a likely safe region to fix the short perturbative scale. 
We show in Fig.~\ref{fig:B_renorm_MSbar} the renormalized $\MSbar$ 
matrix elements computed in  the ratio and the RI-short schemes
at $P^z\simeq 1.7$ GeV. The imaginary part of the matrix elements 
are in full agreement, the same happening in the the real part, for 
large $z$ values. In the small $z$ region of the real part there 
is a tendency to discrepancy, as already observed in the intermediate schemes
shown in Fig.~\ref{fig:B_renorm}, although they agree within one sigma error.

%

\section{Conclusion}
\label{section:conclusions}

In this work, we have addressed the non-perturbative renormalization
of the asymmetric staple-shaped quark bilinear operators. We used symmetry
arguments to restrict the possible mixing between the operators,
and we found that mixing is allowed within sets
of four operators. We then employed the RI/MOM scheme to estimate
the importance of mixing coming from the non-diagonal terms, and we found 
that they can be neglected up to transverse separations of at least $6a$. 
This result justifies the use of multiplicative renormalization for the
asymmetric staple-shaped operator. Based on this conclusion, we 
computed the renormalization factors in the SDR scheme, and in the 
RI-short scheme, where the renormalization factors are computed at
short distances. We found that both schemes are nearly equivalent over
all the $|z|$ region considered. Finally, we converted both schemes to
the $\MSbar$ scheme and presented the corresponding results for 
the renormalized beam function. Given our previous computation of the
Collins-Soper kernel and of the soft function~\cite{Li:2021wvl}, the 
next natural step will be to compute the TMDPDFs themselves over a 
wide range of $b$, and we plan to present results for them in the near future. 

\begin{acknowledgments}
	A.S and F.S are funded by the
	NSFC and the Deutsche Forschungsgemeinschaft 
    (DFG, German Research Foundation) through the funds 
    provided to the Sino-German Collaborative Research Center
	TRR110 “Symmetries and the Emergence of Structure in
	QCD” (NSFC Grant No. 12070131001, DFG Project-ID
	196253076 - TRR 110). 
    K.C. is supported by the National 
    Science Centre (Poland) grant SONATA BIS No.2016/22/E/ST2/00013 
    and OPUS grant No. 2021/43/13/ST2/00497. 
    M.C. acknowledges financial support by the U.S. Department of
	Energy, Office of Nuclear Physics, Early Career Award
	under Grant No. DE-SC0020405. 
    C.A. and J.T. acknowledge partial funding from the project 
    "Unraveling the 3D Parton structure of the nucleon with lattice QCD” 
    (contract id EXCELLENCE/0421/0043)  co-financed by the European Regional 
    Development Fund and the Republic of Cyprus through the Research and 
    Innovation Foundation. 
    G.S. acknowledges funding from the European High-Performance 
    Computing Joint Undertaking (JU) under grant agreement No. 951732 
    and from the European Regional Development Fund and the Republic of 
    Cyprus through the Research and Innovation Foundation under contract 
    No. EXCELLENCE/0421/0025. 
    X.F. is supported in part by NSFC of China under Grant
	No. 11775002 and National Key Research and Development 
    Program of China under Contracts No.2020YFA0406400. X.F and C.L. 
    are supported in part by NSFC of China under Grant No. 12070131001. 
    C.L. is also supported in part by NSFC of China under Grant No. 11935017, 
    No. 12293060, No. 12293062 and No. 12293063.
    This work used computational resources from the John von Neumann-Institute 
    for Computing on the Juwels booster system at the research center in Juelich, 
    under the project with id TMDPDF1.
 
\end{acknowledgments}

\bibliography{refs}

\appendix

\section{Symmetry properties and mixing sets of operators}

As mentioned in Section~\ref{section:operator_mixing}, 
we analyze the symmetry properties of the operators using generalized parity 
and time reversal transformations with discrete flavor rotation, 
as well as  charge conjugation, for the fermion fields in the twisted mass basis, 
$\chi(x_\alpha,\textbf{x})$. The fermion fields in the physical basis, $\psi$ are obtained
through the rotation
\begin{equation}
    \psi = \exp{\left(i\omega\gamma_5\tau^3/2\right)}\chi, \hspace{1cm} 
    \bar{\psi} = \bar{\chi}\exp{\left(i\omega\gamma_5\tau^3/2\right)}.
\end{equation}
The standard parity transformation $\cal P$ is
defined as
\begin{equation}
    \mathcal{P}:
        \begin{cases}
            U(x_0,\vec{x};0) \to U(x_0,-\vec{x};0),\\
            U(x_0,\vec{x};k \in \{1,2,3\}) \to U^{-1}(x_0,-\vec{x}-a\hat{k};k \in \{1,2,3\}),\\
            \chi(x_0,\vec{x}) \to \gamma_0 \chi (x_0,-\vec{x}),\\
            \bar{\chi}(x_0,\vec{x}) \to \bar{\chi}(x_0,-\vec{x}) \gamma_0,
        \end{cases}
\end{equation}
with $U(x_\alpha,\textbf{x};\alpha)$ the gauge link defined
in some direction $\alpha$.
In the twisted basis, this is only a symmetry under a discrete flavour rotation
${\cal F}^{1,2}$ \cite{Shindler:2007vp},
\begin{equation}
    {\cal F}^{1,2}:\begin{cases}
        \chi(x) \to i \tau^{1,2} \chi(x)\\
        \bar{\chi}(x) \to -i \tau^{1,2} \bar{\chi}(x).
    \end{cases}
\end{equation}
The generalized 
parity in the $\alpha$-direction combined with discrete flavor
rotation~\cite{Shindler:2007vp} is then given by
\begin{equation}
        \mathcal{P}_{F \alpha}^{1,2}:
        \begin{cases}
            U(x_\alpha,\textbf{x};\alpha) \to U(x_\alpha,-\textbf{x};\alpha),\\
            U(x_\alpha,\textbf{x};\beta \neq \alpha) \to U^{-1}(x_\alpha,-\textbf{x}-a\hat{\beta};\beta),\\
            \chi(x_\alpha,\textbf{x}) \to i \gamma_\alpha \tau_{1,2} \chi (x_\alpha,-\textbf{x}),\\
            \bar{\chi}(x_\alpha,\textbf{x}) \to -i \bar{\chi}(x_\alpha,-\textbf{x})\tau_{1,2}\gamma_\alpha,
        \end{cases}
        \label{Psymmetry}
\end{equation}
 where the ``standard'' parity is the one with $\alpha=0$, 
 and the $3$-vector $\textbf{x}$ is what remains from the 
 $4$-vector $x$ after removing $x_\alpha$. $\tau_{1,2}$
are the Pauli spin matrices in flavor space. Similarly, the generalized time 
reversal combined with discrete flavor rotation is given by
\begin{equation}
        \mathcal{T}_{F \alpha}^{1,2}:
        \begin{cases}
            U(x_\alpha,\textbf{x};\alpha) \to U^{-1}(-x_\alpha-a,\textbf{x};\alpha),\\
            U(x_\alpha,\textbf{x};\beta \neq \alpha) \to U(-x_\alpha,\textbf{x};\beta),\\
            \chi(x_\alpha,\textbf{x}) \to i \gamma_\alpha \gamma_5 \tau_{1,2} \chi(-x_\alpha,\textbf{x}),\\
            \bar{\chi}(x_\alpha,\textbf{x}) \to -i \bar{\chi}(-x_\alpha,\textbf{x})\tau_{1,2}\gamma_5\gamma_\alpha.
        \end{cases}
        \label{Tsymmetry}
\end{equation}
 Finally, the charge conjugation transformation is given by
\begin{equation}
        \mathcal{C}:
        \begin{cases}
            U(x) \to (U^\dagger(x))^T,\\
            \chi(x) \to C^{-1}\bar{\chi}(x)^T,\\
            \bar{\chi}(x) \to -\chi(x)^T C.
        \end{cases}
        \label{Csymmetry}
\end{equation}
The mixing properties of the staple-shaped operators is
determined from symmetry arguments. Specifically, we observe the 
transformations of the objects $(ijkl)_c$ 
defined by Eq.~(\ref{combinations}). 
This gives rise to 4 relevant combinations with definite symmetry properties: $(++++)_c, (+-+-)_c, (++--)_c$, and $(+--+)_c$. 
In \tab{tab:general_symmetry}, we summarize the symmetries of 
the 16 Dirac structures, where the four signs under any 
transformation denote the symmetry properties of the four relevant combinations, see the caption for more details.

\begin{table}[h]
    \centering
    \begin{tabular}{|c|c|c|c|c|c|c|c|c|}
        \hline
         & $\gamma_0$ & $\gamma_1$ & $\gamma_2$ & $\gamma_3$ 
         & $\mathrm{1}$ & $\gamma_5$ & $\gamma_5\gamma_0$ 
         & $\gamma_5\gamma_1$ \\
        \hline
        ${\cal P}_{F0}^{1,2}$ & $-+-+$ & $+-+-$ & $+-+-$ & $+-+-$ & $-+-+$ 
         & $+-+-$ & $+-+-$ & $-+-+$ \\
        ${\cal P}_{F1}^{1,2}$ & $+-+-$ & $-+-+$ & $+-+-$ & $+-+-$ & $-+-+$ 
         & $+-+-$ & $-+-+$ & $+-+-$ \\
        ${\cal P}_{F2}^{1,2}$ & $++--$ & $++--$ & $--++$ & $++--$ & $--++$
         & $++--$ & $--++$ & $--++$ \\
        ${\cal P}_{F3}^{1,2}$ & $+--+$ & $+--+$ & $+--+$ & $-++-$ & $-++-$ 
         & $+--+$ & $-++-$ & $-++-$ \\
        \hline
        ${\cal T}_{F0}^{1,2}$ & $++++$ & $----$ & $----$ & $----$ & $----$
         & $++++$ & $----$ & $++++$ \\
        ${\cal T}_{F1}^{1,2}$ & $----$ & $++++$ & $----$ & $----$ & $----$
         & $++++$ & $++++$ & $----$ \\
        ${\cal T}_{F2}^{1,2}$ & $-++-$ & $-++-$ & $+--+$ & $-++-$ & $-++-$
         & $+--+$ & $+--+$ & $+--+$ \\
        ${\cal T}_{F3}^{1,2}$ & $--++$ & $--++$ & $--++$ & $++--$ & $--++$
         & $++--$ & $++--$ & $++--$ \\
        \hline
        $C$ & $c$ & $c$ & $c$ & $c$ & $-c$ & $-c$ & $-c$ & $-c$ \\
        \hline
    \end{tabular}
    \newline
    \vspace{1cm}
    \newline
    \begin{tabular}{|c|c|c|c|c|c|c|c|c|}
        \hline
         & $\gamma_5\gamma_2$ & $\gamma_5\gamma_3$ 
         & $\gamma_0\gamma_1$ & $\gamma_0\gamma_2$ & $\gamma_0\gamma_3$ 
         & $\gamma_1\gamma_2$ & $\gamma_1\gamma_3$ & $\gamma_2\gamma_3$ \\
        \hline
        ${\cal P}_{F0}^{1,2}$ & $-+-+$ & $-+-+$ & $+-+-$ & $+-+-$ 
         & $+-+-$ & $-+-+$ & $-+-+$ & $-+-+$ \\
        ${\cal P}_{F1}^{1,2}$ & $-+-+$ & $-+-+$ & $+-+-$ & $-+-+$ 
         & $-+-+$ & $+-+-$ & $+-+-$ & $-+-+$ \\
        ${\cal P}_{F2}^{1,2}$ & $++--$ & $--++$ & $--++$ & $++--$
         & $--++$ & $++--$ & $--++$ & $++--$ \\
        ${\cal P}_{F3}^{1,2}$ & $-++-$ & $+--+$ & $-++-$ & $-++-$
         & $+--+$ & $-++-$ & $+--+$ & $+--+$ \\
        \hline
        ${\cal T}_{F0}^{1,2}$ & $++++$ & $++++$ & $++++$ & $++++$
         & $++++$ & $----$ & $----$ & $----$ \\
        ${\cal T}_{F1}^{1,2}$ & $++++$ & $++++$ & $++++$ & $----$
         & $----$ & $++++$ & $++++$ & $----$ \\
        ${\cal T}_{F2}^{1,2}$ & $-++-$ & $+--+$ & $-++-$ & $+--+$
         & $-++-$ & $+--+$ & $-++-$ & $+--+$ \\
        ${\cal T}_{F3}^{1,2}$ & $++--$ & $--++$ & $--++$ & $--++$
         & $++--$ & $--++$ & $++--$ & $++--$ \\
        \hline
        $C$ & $-c$ & $-c$ & $c$ & $c$ & $c$ & $c$ & $c$ & $c$ \\
        \hline
    \end{tabular}
    \caption{General symmetry properties of all possible Dirac structures. The first sign of each entry reflects the symmetry properties ($+/-$ -- symmetric/antisymmetric) of the combination $(++++)_c$ (or its equivalent $(----)_c$), the second sign concerns $(+-+-)_c$ (or $(-+-+)_c$),
    the third $(++--)_c$ (or $(--++)_c$) and the fourth $(+--+)_c$ (or $(-++-)_c$). The last row indicates the symmetry properties with respect to charge conjugation that depend on the sign $c$ and are common to all combinations. The combinations that mix have the same signs in a given column (see e.g.\ the first column of $\gamma_0$, the second column of $\gamma_0\gamma_2$, the third column of $\gamma_5\gamma_1$ and the fourth column of $\gamma_0\gamma_3$, representing the quadruple $\left\{\gamma_0,\gamma_0\gamma_2,\gamma_0\gamma_3,\gamma_5\gamma_1\right\}$ of operators that mix relevant to this work; note that the mixing with $\gamma_5\gamma_1$ involves the combination $(++--)_{-c}$ with opposite relative sign in front of the charge-conjugated operators).}
    \label{tab:general_symmetry}
\end{table}

Based on the symmetry properties,  we provide all the mixing sets of asymmetric ($z\neq0$) staple-shaped operators with different Dirac structures $\Gamma$, as dictated by the generalized C, P, T symmetries of Eqs. (\ref{Psymmetry} -- \ref{Csymmetry}), in 
explicit form:

\begin{itemize}
\item $\{1, \gamma_2, \gamma_3, \gamma_2 \gamma_3\}$,
\item $\{\gamma_5, \gamma_5 \gamma_2, \gamma_5 \gamma_3, \gamma_0 \gamma_1\}$,
\item $\{\gamma_0, \gamma_0 \gamma_2, \gamma_0 \gamma_3, \gamma_5 \gamma_1\}$,
\item $\{\gamma_1, \gamma_1 \gamma_2, \gamma_1 \gamma_3, \gamma_5 \gamma_0\}$.
\end{itemize}

In the case of symmetric staple-shaped operators ($z=0$), the mixing sets are reduced to:
\begin{itemize}
\item $\{\gamma_2, \gamma_3, \gamma_2 \gamma_3\}$,
\item $\{\gamma_5, \gamma_5 \gamma_2, \gamma_5 \gamma_3\}$,
\item $\{\gamma_0, \gamma_0 \gamma_2, \gamma_0 \gamma_3\}$,
\item $\{\gamma_1, \gamma_1 \gamma_2, \gamma_1 \gamma_3\}$,
\end{itemize}
while the remaining operators, involving the Dirac structures $1, \gamma_0 \gamma_1, \gamma_5 \gamma_1, \gamma_5 \gamma_0$ are multiplicatively 
renormalizable.

\end{document}